\documentclass{emulateapj}
\usepackage{epstopdf}
\usepackage{subfigure}
\usepackage{rotating}
\usepackage{hyperref}
\usepackage{amsmath}
\usepackage{mathtools}

\newcommand{\feh}{\hbox{$ [{\rm Fe}/{\rm H}]$ }} 
\def\Sec{${}^{\prime\prime}$\llap{.}}

\slugcomment{accepted to AJ}

\shorttitle{NGC$\,$2808 variables}
\shortauthors{Kunder et~al.}

\begin{document}

\title{Variable stars in the Globular Cluster NGC$\,$2808}

\author{Andrea Kunder\altaffilmark{1},
Peter B. Stetson\altaffilmark{2}, 
M\'{a}rcio Catelan\altaffilmark{3,4},
Alistair R. Walker\altaffilmark{1},
and
P\'{i}a Amigo\altaffilmark{3,4},  
}

\altaffiltext{1}{Cerro Tololo Inter-American Observatory, Casilla 603, La Serena, Chile}
\affil{E-mail: akunder@ctio.noao.edu}
\altaffiltext{2}{ Dominion Astrophysical Observatory, Herzberg Institute of Astrophysics, National Research Council, Victoria BC, Canada}
\altaffiltext{3}{Pontificia Universidad Cat\'olica de Chile, Departamento de Astronom\'\i a y Astrof\'\i sica, Av. Vicu\~{n}a Mackenna 4860, 782-0436 Macul, Santiago, Chile; e-mail: mcatelan@astro.puc.cl}  
\altaffiltext{4}{The Milky Way Millennium Nucleus, Av. Vicu\~{n}a Mackenna 4860, 782-0436 Macul, Santiago, Chile} 

\begin{abstract}
The first calibrated broadband {\it BVI\/} time-series photometry is presented for the variable stars
in NGC$\,$2808, with observations spanning a range of twenty-eight years.  
We have also redetermined the variability types and periods for the variable stars
identified previously by Corwin et~al, revising the number of probable fundamental-mode 
RR Lyrae variables (RR0) to 11 and the number of first-overtone variables (RR1) 
to five.  Our observations were insufficient to discern the nature of the
previously identified RR1 star, V24, and the tentatively identified
RR1 star, V13.  These two variables are $\sim$0.8 mag brighter than the RR Lyrae variables,
appear to have somewhat erratic period and/or luminosity changes, and lie inside the 
RR Lyrae instability strip.  
Curiously, all but one of the RR Lyrae stars studied in this relatively metal-rich cluster
exhibit the Blazhko phenomenon, an effect thought to occur with higher frequency in 
metal-poor environments.  The mean periods of the RR0 and RR1 variables are 
$\langle P \rangle_{RR0}=$0.56$\pm$0.01 d and $\langle P \rangle_{RR1}=$0.30$\pm$0.02 d, respectively, 
supporting an Oosterhoff I classification of the cluster.  On the other hand, the number ratio 
of RR1- to RR0-type variables is high, though not unprecedented, for an Oosterhoff I 
cluster.  The RR Lyrae variables have no period shifts at a given amplitude as 
compared to the M3 variables, making it unlikely that these variables are He-enhanced.  
Using the recent recalibration of the RR Lyrae luminosity scale by Catelan \& Cort\'{e}s, 
a mean distance modulus of 
$\rm (m-M)_V$= 15.57$\pm$0.13 mag for NGC$\,$2808 is obtained, in good agreement 
with that determined here from its type II Cepheid and SX Phoenicis population.  
Our data have also allowed the discovery of two new candidate SX Phoenicis stars and an 
eclipsing binary in the blue straggler region of the NGC$\,$2808 color-magnitude diagram.  
\end{abstract}


\section{Introduction}
NGC$\,$2808 is one of the most luminous globular clusters (GCs) in the Galaxy 
($M_V$=$-$9.39) and also one of the most magnificent, especially in terms of its 
color-magnitude diagram (CMD).  Not only is its main sequence (MS) a fusion of three 
distinct sequences interpreted as successive episodes of star formation \citep{dantona05, piotto07}, 
but its horizontal branch (HB) is multimodal with a greatly extended hot blue tail 
that covers a range of $\sim$5 mag, in $V$, below the mean level of the instability strip 
(IS) \citep{bedin00, dalessandro11}.  A multimodal HB is not easily explained; GCs with 
red HB stars cooler than the RR Lyrae gap are generally associated with relatively high
metallicities, while those with blue HB stars tend to be metal-poor systems.
But multimodal HBs do not fit into this common paradigm, and therefore such clusters
may help provide a key in a more general understanding of the HBs of GCs
\citep[e.g.,][and references therein]{catelan98a,catelan08,catelan09a}.  

The three MS branches in NGC$\,$2808 are thought to be associated with structures in its HB:  it
has been put forth by \citet{dantona05} that the three MS populations all have
different helium abundances leading to three different HB populations.  
According to their models, the RR Lyrae variables (RRLs) could be either 
the bluest extension of the red clump population, which has a normal He, 
or they could be the reddest part of the blue clump, which contains 
stars slightly helium-enhanced (0.26$<$ $Y$ $<$0.29).  Although the 
existence of the high-helium ($Y$$\sim$ 0.40) population as well as the 
normal-helium population seems robust \citep{dantona05, bragaglia10, pasquini11}, 
the intermediate-helium population is inferred only from the HB morphology.  
In this sense, the RRLs are particularly valuable as potential support for the 
existence of the intermediate-helium population.  Similarly, models from 
\citet{lee05} and \citet{dalessandro11} predict a normal He for the red HB, 
an enhanced He for the blue HB, but by and large do not discuss the instability strip.

Here an observational analysis of one of the HB components in NGC$\,$2808, 
the RR Lyrae instability strip, is presented.  This component is largely unexplored, 
especially because RRLs were not even thought to be a major constituent of 
NGC$\,$2808 until \citet{corwin04} discovered a number of these variables,
raising the number of candidate RRL to 18.  
The \citet{corwin04} variables are all located in its crowded core, so 
placing the light curves on a photometrically calibrated scale was not
possible then.  

However, investigating the helium enhancement of RR Lyrae variables requires a 
knowledge of their luminosities \citep[e.g.,][]{asano68, vanalbada71,sweigart87,
caceres08, dalessandro11}.  This is because their mean luminosity
is a function of He.  Calibrated amplitudes are also important because these
are correlated to their $T_{eff}$ , which in turn can be compared to their period
to understand potential ``period shifts" \citep[e.g.,][]{carney92, catelan98, desantis01}.  
The period shift is usually made between
the RRLs in different globular clusters \citep{sandage81, sandagea81,
carney92, sandage93}, and can be used as an indicator for
$\rm [Fe/H]$, luminosity, or Oosterhoff (Oo) type \citep[e.g.,][]{cacciari05}. In particular, if
the enhancement of He is present in the RRLs, large period shifts are clearly 
expected \citep[see, e.g.][]{sweigart87, catelan98b, marconi11, kunder12}.

In this paper new and archival {\it BVI\/} photometry of NGC$\,$2808 is presented, which 
has allowed the first calibrated light curves of the RR Lyrae variables.
The discovery of additional variable stars and the re-examination of the currently known 
variables in the cluster are also reported.
Constraints from the ancient RRLs about the cluster's formation history are put forward.
\section{Observations}
The observational material for this study consists of 2,475
individual CCD images from 38 observing runs.  These images
are contained within a data archive maintained by author
PBS, and include --- among others --- the images
employed by \citet{corwin04}, as well as some images obtained
specifically for this project:  five nights of observations in
January 2010, two nights in April 2010, and four nights in January
2011, all carried out on the SMARTS 1m telescope on Cerro Tololo. 
Summary details of all 38 observing runs are given in
Table~\ref{obs}.  Considering all these images together, the
median seeing for our observations was 1.09 arcseconds; the 25th
and 75th percentiles were 0.88 and 1.35 arcseconds; the 10th and
90th percentiles were 0.74 and 1.69 arcseconds.  

The photometric reductions were all carried out 
using standard DAOPHOT/ALLFRAME procedures \citep{ste87, ste94} 
to perform profile-fitting photometry, which was then referred to a
system of synthetic-aperture photometry by the method of growth-curve
analysis \citep{ste90}.  Calibration of these instrumental data
to the photometric system of \citet[][see also Landolt 1973, 1983]{landolt92} 
was carried out as described by \citet{ste00, ste05}.  If we
define a ``dataset" as the accumulated observations from one CCD
chip on one night with photometric observing conditions, or one
chip on one or more consecutive nights with non-photometric
conditions, the data for NGC$\,$2808 were contained within 90
different datasets, each of which was individually calibrated to
the Landolt system.  Of these 90 datasets, 61 were obtained and
calibrated as ``photometric,'' meaning that photometric zero
points, color transformations, and extinction corrections were
derived from all standard fields observed during the course of
that night, and applied to the NGC$\,$2808 observations.  The other 29
datasets were reduced as ``non-photometric"; in this case, color
transformations were derived from all the standard fields
observed, but the photometric zero point of each individual CCD
image of NGC$\,$2808 was derived from local photometric standards
contained within the image itself.  

The different cameras employed projected to different areas on the
sky, and of course the telescope pointings differed among the
various exposures.  The EMMI MIT/LL, FORS, SOI and WFI magers, in
particular, consist of mosaics of non-overlapping CCD detectors. 
Therefore, although we have 2,475 images, clearly no individual
star appears in all those images.  In fact, no star appeared in
more than 73 $U$-band images, 672 $B$ images, 564 $V$ images, or
267 $I$ images.  The {\it median\/} number of observations for one
of our stars is 16 in $U$, 143 in $B$, 248 in $V$, and 147 in $I$. 
Since most pointings were centered on or near the cluster, member
stars typically have more observations than field stars lying far
from the cluster center.

There were insufficient $R$-band images to define local standards
in the NGC$\,$2808 field, so we have not calibrated the $R$ data. 
Furthermore, during the ``dan0210'' observing run (number 19 in
Table~1) there were some 38 images apparently obtained in a
long-wavelength band, most likely the $Z$-band; they were incorrectly
identified as ``bessell B'' in the FITS image headers, but from the
magnitude differences observed between red/blue pairs of stars,
the filter evidently had a longer effective wavelength than $I$. 
These $Z$- and $R$-band images were included in the ALLFRAME
analysis for the additional information they give on the
completeness of the star list and the precision of the astrometry,
but we make no further use of them here.  

It should also be noted that calibration of $U$-band data is
notoriously problematic because significantly different
prescriptions for the $U$ filter are employed in different filter
sets, the quantum efficiency of many CCDs is highly
wavelength-dependent within the $U$ bandpass, and the form of the
bandpass itself is further modified by the wavelength-dependent
transparency of the terrestrial atmosphere.  Thus, while we have
obtained $U$-band photometry for NGC$\,$2808 and have calibrated
it to the Landolt system to the best of our ability, we do not
employ it here for any scientifically critical inferences.  

The astrometric calibration is tied to the U.S. Naval Observatory (USNO) 
A2.0 Astrometric Reference Catalog.  Positions of stars encompassing our 
target field were downloaded from the USNO interface of the Canadian 
Astronomy Data Centre (CADC).  From the CADC we also downloaded 
all available images of our field from the Digitized Sky Survey; in the case 
of NGC$\,$2808, scans of five plates were available:  one IIIaJ+GG395 ($\sim B$), 
one IIaD+GG495 ($\approx V$), two IIIaF+OG590 ($\approx R$), and one 
IVN+RG715 ($\sim I$) plate.  Positions and magnitude indices of star-like 
objects in the digitized photographic images were obtained using the software 
of \citet{ste79}.  These positions were referred to the coordinate system of 
the USNO A2.0 catalog employing 20-constant cubic polynomial transformations 
based on 10,000 to 22,000 stars per plate, and averaged.  Then the positions of 
stars in the CCD images were all transformed to this average coordinate system 
via cubic polynomials, and averaged.   We find that, compared to the weighted 
average of all measured positions, the USNO A2.0 catalog has root-mean-square 
positional residuals of 0\Sec22 in each coordinate; each of the five digitized 
plates had r.m.s. residuals ranging from 0\Sec16 to 0\Sec23; and the CCD 
positions had r.m.s.\ residuals of 0\Sec06.  Since these transformations were 
based on more than 20,000 stars, we believe that it is safe to say that our 
positions are on the USNO A2.0 astrometric system to within well 
under 0\Sec1.  If the USNO system itself contains systematic errors larger than this 
relative to some other ``true'' astrometric system, then our positions have inherited 
those systematic errors.  We believe that within that overall system, our relative 
star-to-star positions are internally precise to better than 0\Sec03 depending, of course, 
on the signal-to-noise ratio of the individual stellar detections. 
 
\clearpage

\begin{table}
\caption{NGC$\,$2808 Observations}
\label{obs}
\begin{tabular}{lllcccclcc} \hline
Run ID & Dates & Telescope/Camera/Detector  & U & B & V & R & I & Note \\ \hline
\hline
 1 ct83      & 1983 Feb 12    & CTIO 4.0m RCA                  & -- &   2 &  2 & -- & -- \\
 2 ct84      & 1984 Mar 07    & CTIO 4.0m RCA                  & -- &   1 &  1 & -- & -- \\
 3 ct85      & 1985 Apr 18    & CTIO 4.0m RCA                  & -- &   1 &  1 & -- & -- \\
 4 pab       & 1987 Jan 23-24 & CTIO 0.9m RCA                  & -- &  12 & 12 & -- & -- \\
 5 aat       & 1991 Jan 13    & AAT 3.9m CCD\_1                & -- &   6 & -- & 15 & -- \\
 6 emmi3     & 1994 Dec 26    & ESO NTT 3.6m EMMI TK2048EB     & -- &  -- &  4 & -- &  3 \\
 7 emmi2     & 1995 Mar 08-10 & ESO NTT 3.6m EMMI TK2048EB     & -- &  -- & 18 & -- & 19 \\
 8 zingle    & 1996 Apr 16-19 & CTIO 0.9m Tek2K\_3             & -- &  -- & 18 & -- & 16 \\
 9 apr97     & 1997 Apr 12    & ESO Dutch 0.91m Tektronix $33$ & -- &  -- &  6 & -- &  6 \\
10 bond5     & 1997 Jun 02    & CTIO 0.9m Tek2K\_3             &  2 &   2 &  2 & -- &  2 \\
11 bond3     & 1997 Nov 09-11 & CTIO 0.9m Tek2K\_3             &  3 &   3 &  3 & -- &  3 \\
12 dec97     & 1997 Dec 26-27 & ESO Dutch 0.91m Tektronix $33$ & -- &  -- & 14 & -- & 14 \\
13 rolly     & 1998 Jan 26-28 & ESO Danish 1.54m DFOSC Loral   &  6 &  20 & 17 & -- & 10 \\
14 bond6     & 1998 Apr 17-19 & CTIO 0.9m Tek2K\_3             &  2 &   2 &  2 & -- &  2 \\
15 bond4     & 1999 Jun 13-16 & CTIO 0.9m Tek2K\_3             &  4 &   2 &  2 & -- &  2 \\
16 wfi11     & 1999 Dec 06-07 & ESO MP 2.2m WFI                &  8 &  10 &  9 & -- & -- & $^a$ \\
17 bond7     & 2001 Mar 25    & CTIO 0.9m Tek2K\_3             &  1 &   1 &  1 & -- &  1 \\
18 wfi6      & 2002 Feb 19-21 & ESO 2.2m MP WFI                & -- &   4 &  4 & -- &  6 & $^a$ \\
19 dan0210   & 2002 Oct 15-16 & ESO Danish 1.54m DFOSC MAT/EEV & -- &  -- & -- & -- & 36 \\
20 dan0212   & 2002 Dec 12-15 & ESO Danish 1.54m DFOSC MAT/EEV & -- &  76 & 72 & -- & -- \\
21 dan0302   & 2003 Feb 19-20 & ESO Danish 1.54m DFOSC MAT/EEV & -- &  83 & 81 & -- & -- \\
22 emmi0305  & 2003 May 10-11 & ESO NTT 3.6m EMMI emred        & -- &  -- & 74 & -- & -- \\
23 fors3     & 2005 Feb 16-17 & ESO VLT 8.0m FORS2             & -- &  13 & 19 & -- & -- & $^b$ \\
24 dan0504   & 2005 Apr 29    & ESO Danish 1.54m DFOSC EEV     & -- &  23 & 21 & -- & -- \\
25 dan0505   & 2005 May 12-13 & ESO Danish 1.54m DFOSC EEV     & -- &  32 & 33 & -- & -- \\
26 fors20605 & 2006 May 29    & ESO VLT 8.0m FORS2 ccdf        & -- &  -- &  1 &  3 &  1 & $^b$ \\
27 soar0906  & 2009 Feb 27    & SOAR 4.1m SOI                  & 10 &   9 &  9 & -- & 13 & $^b$ \\
28 efosc0904 & 2009 Apr 16    & ESO NTT 3.6m EFOSC LORAL       & -- & 240 & -- & -- & -- \\
29 efosc09   & 2009 Apr 19-26 & ESO NTT 3.6m EFOSC LORAL       & -- &  68 &  7 & -- & -- \\
30 pia       & 2010 Jan 10-15 & CTIO 1.0m Y4KCam               & -- &  15 & 15 & -- & 15 \\
31 soar10a   & 2010 Jan 14    & SOAR 4.1m SOI                  &  3 &   5 &  6 & -- &  4 & $^b$ \\
32 soar10bc  & 2010 Jan 20-21 & SOAR 4.1m SOI                  & 11 &  11 &  7 & -- &  9 & $^b$ \\
33 soar10d   & 2010 Feb 08    & SOAR 4.1m SOI                  &  7 &   7 &  7 & -- & 14 & $^b$ \\
34 soar10j   & 2010 Oct 14    & SOAR 4.1m SOI                  & -- &  30 & 30 & -- & 28 & $^b$ \\
35 andrea1   & 2010 Apr 27-28 & CTIO 1.0m Y4KCam               & -- &  -- & 18 & 17 & -- \\
36 andrea2   & 2010 Jun 27    & CTIO 1.0m Y4KCam               & -- &  -- &  4 &  3 & -- \\
37 Y1101     & 2011 Jan 13-18 & CTIO 1.0m Y4KCam               & -- &  46 & 73 & 66 & 50 \\
38 Y1104     & 2011 Apr 14-18 & CTIO 1.0m Y4KCam               & -- &  -- & 69 &  1 & 57 \\
\hline
\hline
\end{tabular}
\\
$^a$ eight individual CCD chips in the camera; the reported number of CCD images is the number of exposures x 8 \\
$^b$ two individual CCD chips in the camera \\
\\
Notes: \\ 
 1. Observers J.~Hesser \& R. McClure
 2. Observers J.~Hesser \& R. McClure
 3. Observers J.~Hesser \& R. McClure
 4. Observer P.~Bergbusch
 5. Observers Da~Costa \& Norris
 6. Observer Testa, Project ID 054.E-0404
 7. Observer Zaggia, Project ID 054.E-0337
 8. Observer R.~Zingle
 9. Observer A.~Rosenberg
10. Observer H.~E.~Bond
11. Observer H.~E.~Bond
12. Observer A.~Rosenberg
13. PI L.~Bedin
14. Observer H.~E.~Bond
15. Observer H.~E.~Bond
16. Observer A.~Recio Blanco, Program ID 064.L-0255
17. Observer H.~E.~Bond
18. Program ID 68.D-0265(A)
19. Observer T.~M.~Corwin, M.~Catelan, \& J.~M.~Fern\'{a}ndez, Program ID 02/8232
20. Observer J.~Borissova \& M.~Catelan, Program ID 02/8232
21. Observer T.~M.~Corwin, M.~Catelan, Program ID 02/8232
22. Program ID 071.D-0609(A)
23. Program ID 074.D-0187(B)
24. Observer R.~Salinas
25. Observer R.~Salinas
26. Program ID 077.D-0775(A)
27. Observer M.~Smith, Proposal ID 2008B-0482
28. Program ID 083.D-0833(A)
29. Program ID 083.D-0544(A)
30. Observers C.~Gatica \& A.~Brito
31. Observer ``MSU", Proposal ID 2009B-0340
32. Proposal ID 2009A-0414
33. Observer Smith, Proposal ID 2010A-0312
34. Observers S.~Zepf \& M.~Peacock
35. Observers A.~Kunder, J.~Subasavage \& J.~Seron
36. Observer A.~Kunder
37. Observers A.~Kunder \& S.~A.~Stubbs
38. Observer A.~Kunder 
  \end{table}
\clearpage

 
\begin{table}[ph!]
\begin{scriptsize}
\centering
\caption{Variable stars detected in NGC$\,$2808}
\label{lcpars}
\begin{tabular}{p{0.2in}p{0.55in}p{0.58in}p{0.43in}p{0.28in}p{0.28in}p{0.28in}p{0.28in}p{0.28in}p{0.28in}p{0.2in}p{0.2in}p{0.2in}p{0.45in}p{0.45in}} \\ \hline
Name & R.A. (J2000.0) & Decl. (J2000.0) & Period (d) & $(B)_{mag}$ & $(V)_{mag}$ & $(I)_{mag}$ & $\hbox{\it $<$B$>$\/} $ & $\hbox{\it $<$V$>$\/} $  & $\hbox{\it $<$I$>$\/} $ & $A_B$ & $A_V$ & $A_I$ & Type & Comment \\ 
\hline
V1 & 09 12 20.22 & $-$64 52 20.5 & -- & 14.8 & 13.4 & 11.2 & -- & -- & -- & -- & -- & -- & LPV &  \\
V2 & 09 11 55.66 & $-$64 51 10.5 & -- & 15.92 & 14.64 & 13.23 & -- & -- & -- & -- & -- & -- & NV &  \\
V3 & 09 12 08.38 & $-$64 51 47.3 & -- & 16.83 & 15.82 & 14.64 & -- & -- & -- & -- & -- & -- & NV? & large W/S index, poss. P=80 d \\
V4 & 09 11 33.77 & $-$64 51 43.8 & -- & 19.00 & 18.09 & 16.96 & --& --& --& --& --& -- & NV &  \\
V5 & 09 12 09.57 & $-$64 51 52.4 & -- & 16.62 & 16.49 & 16.25 & -- & -- & -- & -- & -- & -- &  NV &  \\
V6 & 09 12 30.08 & $-$64 56 37.9 & 0.538968 & 16.75 & 16.31 & 15.62 & 16.74 & 16.29 & 15.60 & 1.08 & 0.75 & 0.71 & RR0 & Blazhko \\
V7 & 09 13 13.13 & $-$64 50 46.5 & -- & 15.16 & 13.54 & 11.82 & -- & -- & -- & -- & -- & -- & NV &  \\
V8 & 09 12 06.56 & $-$64 53 49.3 & -- & 15.83 & 14.76 & 13.50 & -- & -- & -- & -- & -- & -- & NV &  \\
V9 & 09 11 53.06 & $-$64 54 24.3 & -- & 16.82 & 16.24 & 15.41 & -- & -- & -- & -- & -- & -- & NV &  \\
V10 & 09 11 56.86 & $-$64 53 23.2 & 1.76528 & 15.92 & 15.30 & 14.48 & 15.91 & 15.28 & 14.47 & 1.02 & 0.73 & 0.47 & BL Her &  \\
V11 & 09 12 06.72 & $-$64 52 40.3 & -- & 14.5 & 12.9 & 11.1 & -- & -- & -- & -- & -- & -- & LPV &  \\
V12 & 09 11 56.15 & $-$64 50 09.6  & 0.305776 & 16.62 & 16.26 & 15.72 & 16.63 & 16.27 & 15.71 & 0.58 & 0.45 & 0.28 & RR1 & Blazhko \\
V13 & 09 12 23.35 & $-$64 57 10.6 & 0.173678 & 16.09 & 15.52 & 14.76 &16.26 & 15.59 & 14.79 & 0.18 & 0.18 & 0.16 & W UMa? ellipsoidal binary? & field star \\
V14 & 09 12 13.40 & $-$64 51 04.0 & 0.598903 & 16.75  & 16.27 & 15.53 & 16.76 & 16.26 & 15.53 & 0.98 & 0.73 & 0.45 & RR0 & \\
V15 & 09 12 10.61 & $-$64 51 15.7 & 0.610975 & 16.82 & 16.31 & 15.46 &16.84 & 16.29 & 15.46 & 0.46 & 0.38 & 0.23 & RR0 & Blazhko \\
V16 & 09 12 09.11 & $-$64 52 51.5 & 0.605218 & 16.75 & 16.32 & 15.67 &16.74 & 16.29 & 15.68 & 0.70 & 0.66 & 0.46 & RR0 & Blazhko \\
V17 & 09 12 07.01 & $-$64 51 57.3 & 0.504050:: & -- & -- & -- & -- & -- & -- & -- & -- & -- & RR0? & blend \\ 
V18 & 09 12 06.57 & $-$64 52 17.0 & 0.5851455 & -- & -- & -- & -- & -- & -- & -- & -- & -- & RR0 & Blazhko, slightly blended \\ 
V19 & 09 12 06.26 & $-$64 52 10.7 & 0.5091927:: & -- & -- & -- & -- & -- & -- & -- & -- & -- & RR0 & blend \\
V20 & 09 12 05.32 & $-$64 51 26.5 & 0.287021 & 16.55 & 16.01 & 15.37 & 16.55 & 16.04 & 15.37 & 0.69 & 0.43 & 0.29 &  RR1 & Blazhko \\ 
V21 & 09 12 03.90 & $-$64 51 25.9 & 0.604334 & -- & -- & -- & -- & -- & -- & -- & -- & -- & RR0 & blend \\ 
V22 & 09 12 03.87 & $-$64 51 55.3  & 0.518688 & -- & -- & -- & -- & -- & -- & -- & -- & -- & RR0 & blend \\ 
V23 & 09 12 03.26 & $-$64 52 19.7 & 0.265985 & 16.51 & 16.05 & 15.47 & 16.50 & 16.08 & 15.47 & 0.56 & 0.34 & 0.16 & RR1 & Blazhko \\
V24 & 09 12 02.62 & $-$64 52 10.7 & 0.268199:: & 16.12 & 15.55 & 14.74 & 16.18 & 15.66 & 14.83 & 0.54 & 0.24 & 0.19 & W UMa? & properties uncertain\\  
V25 & 09 11 58.68 & $-$64 51 27.9 & 0.49517:: & -- & -- & -- & -- & -- & -- & -- & -- & -- & RR0 & blend \\
V26 & 09 11 53.09 & $-$64 51 30.7  & 0.365099 & 16.48 & 16.16 & 15.63 & 16.47 & 16.15 & 15.63 & 0.53 & 0.42 & 0.32 & RR1 & Blazhko \\
V27 & 09 11 53.64 & $-$64 52 59.7 & 0.573229 & -- & -- & -- & -- & -- & -- & -- & -- & -- & RR0 & blend \\ 
V28 & 09 11 50.58 & $-$64 51 27.8 & 0.2767642 & 16.58 & 16.25 & 15.74 & 16.56 & 16.25 & 15.76 & 0.64 & 0.53 & 0.36 & RR1 & Blazhko \\
V29 & 09 12 00.53 & $-$64 51 39.2 & 2.22211 & 15.50 & 14.82 & 13.90 & 15.52 & 14.84 & 13.92 & 0.70 & 0.52 & 0.38 & BL Her &  \\
V30 & 09 11 49.37 & $-$64 51 52.6 & 3.25710 & 17.84 & 17.48 & 16.83 & 17.85 & 17.49 & 16.84 & -- & 1.42 & -- & E & detached eclipsing binary \\
V31 & 09 11 57.08 & $-$64 51 29.7 & -- & 15.25 & 13.39 & 11.35 & -- & -- & -- & -- & -- & -- & LPV & LW1 \\
V32 & 09 11 58.71 & $-$64 51 28.6 & -- & 15.19 & 13.72 & 12.18 & -- & -- & -- & -- & -- & -- & LPV & LW2 \\
V33 & 09 11 59.24 & $-$64 52 59.2 & -- & 15.14 & 13.39 & 11.36 & -- & -- & -- & -- & -- & -- & LPV & LW3 \\
V34 & 09 12 01.44 & $-$64 51 37.7 & --  & -- & 13.16 & 11.3 & -- & -- & -- &  -- & -- & -- & LPV & LW4 \\
V35 & 09 12 01.69 & $-$64 50 33.3 & -- & 15.40 & 13.52 & 11.55 & -- & -- & -- & -- & -- & --  & LPV & LW5 \\
V36 & 09 12 02.28 & $-$64 51 18.5 & -- & 15.1 & 13.45 & 11.64 & -- & -- & -- & -- & -- & -- & LPV & LW6 \\
V37 & 09 12 02.69 & $-$64 52 01.9 & -- & 15.1 & 13.7 & 12.1 & -- & -- & -- & -- & -- & -- & LPV & LW7 \\
V38 & 09 12 04.34 & $-$64 51 41.5 & -- & 15.0 & 13.42 & 11.36 & -- & -- & -- & -- & -- & -- & LPV & LW8 \\
V39 & 09 12 06.66 & $-$64 52 18.2 & -- & 15.2 & 13.51 & 11.61 & -- & -- & -- & --  & -- & -- & LPV & LW9 \\
V40 & 09 12 08.51 & $-$64 51 58.4 & -- & 15.15 & 13.44 & 11.64 & -- & -- & -- & -- & -- & -- & LPV & LW10 \\
V41 & 09 12 11.30 & $-$64 52 42.2 & -- & 15.3 & 13.47 & 11.62 & -- & -- & -- &  -- & -- & -- & LPV & LW11 \\
V42 & 09 12 14.47 & $-$64 53 26.7 & -- & 15.14 & 13.40 & 11.51 & -- & -- & -- & -- & -- & -- & LPV & LW12 \\
V43 & 09 12 16.64 & $-$64 52 02.9 & -- & 15.23 & 13.51 & 11.58 & -- & -- & -- & -- & -- & -- & LPV & LW13 \\
V44 & 09 12 18.48 & $-$64 51 30.7 & -- & 15.24 & 13.72 & 12.06 & -- & -- & -- & --  & -- & -- & LPV & LW14 \\
V45 & 09 12 04.02 & $-$64 51 54.0 & -- & -- & 13.46 & -- & -- & -- & -- & -- & -- & -- & LPV & LW15 \\
V46 & 09 11 50.35 & $-$64 51 42.2 & -- & 15.68 & 13.76 & 11.73 & -- & -- & -- & -- & -- & -- & LPV & LW16 \\ 
V47 & 09 12 01.49 & $-$64 50 49.4 & -- & 15.33 & 13.77 & 12.13 & -- & -- & -- & -- & -- & -- & LPV & LW17 \\ 
V48 & 09 12 04.85 & $-$64 51 30.0 & -- & 16.02 & 14.96 & 13.7 & -- & -- & -- & -- & -- & -- & LPV & LW18 \\
V49 & 09 12 05.88 & $-$64 52 08.1 & -- & 15.05 & 13.41 & 11.67 & -- & -- & -- & -- & -- & -- & LPV & LW19 \\
V50 & 09 12 24.79 & $-$64 51 10.4 & -- & 15.26 & 13.55 & 11.72 & -- & -- & -- &  -- & -- & -- & LPV & LW20 \\
\hline
V51 & 09 12 05.52 & $-$64 51 31.1 & 2.10797 & 15.73 & 15.10 & 14.26 & 15.73 & 15.09 & 14.26 & 0.72 & 0.54 & 0.67 & BL Her &  \\
C52 & 09 11 48.52 & $-$64 46 59.5 & 0.05981586 & 18.74 & 18.33 & 17.75 & 18.74 & 18.33 & 17.76 & 0.39 & 0.30 & 0.18 & SX Phe &   \\
C53 & 09 11 29.23 & $-$64 45 23.3 & 0.05181968 & 18.68 & 18.32 & 17.80 & 18.68 & 18.32 & 17.80 & 0.24 & 0.22 & 0.13 & SX Phe & poss. second period of P=0.04011174 \\
C54 & 09 11 40.25 & $-$64 53 12.8 & 0.441618 & 18.70 & 18.29 & 17.73 & 18.69 & 18.29 & 17.73 & 0.30 & 0.28 & 0.26 & EB \\
\hline
\end{tabular}
\end{scriptsize}
\end{table}

\clearpage

\section{Variable Stars in NGC$\,$2808}
Several authors have reported variable stars in NGC$\,$2808.  The first
finding chart of seven variables was published by
\citet{fourcade66}, although five of these are now
thought to be actually non-varying \citep{clement89}.  \citet{alcaino71} identified two
further variables, which were also later determined to be non-varying \citep{clement89}.
The study by \citet{clement89} produced three new variables, bringing the number of
confirmed variables to five.  With the advent of modern CCD observations
and data reduction techniques, \citet{corwin04} announced 18 new
variables close to the crowded cluster core.  Unfortunately, because of the compact nature of this 
cluster, they were unable to place the light curves on a photometrically calibrated scale.  
The first candidate variable stars using the {\it Hubble Space Telescope (HST)} were
found by \citet{dieball05}.  As the {\it HST} observations were taken over four days only,
reliable light curves and periods for the six variable candidates were not obtained.
One of their candidates corresponds with V22, an RRL identified by \citet{corwin04},
and the rest of their candidates are considerably fainter.
Long-period variables (LPVs) were searched for by \citet{lebzelter11}, resulting in the 
determination of periods and amplitudes for 15 LPVs.  

Besides the LPVs, no calibrated CCD time-series photometry has been presented for any 
of the confirmed variables in NGC$\,$2808.  The time-series observations presented
here cover a baseline that allows for the potential to detect variability on time-scales from 
hours to months.  Therefore, we are in a position to re-examine and revise the list of 
variables in the cluster.

The characteristics of the confirmed variables are given in Table~\ref{lcpars}.
Although seven previously identified variables were too blended to derive reliable calibrated 
magnitudes and amplitudes, we were able to obtain periods for these stars from
their instrumental light curves.  The
positions of all the variables, as well as for the constant stars (NV) that have been 
incorrectly identified as variables by others, are also listed for completeness.
We estimate that the astrometry presented is accurate to better than 0.1~arcseconds
(see \S2).  The columns contain 
(1) the name of the variable as given in the 2011 update of  NGC$\,$2808 in the 
\citet{clement01} catalog, 
(2) the right ascension in hours, minutes and seconds (epoch J2000), 
(3) the declination in degrees, arcminutes and arcseconds, 
(4) the period in days, 
(5-7) the magnitude-weighted mean $B$, $V$, and $I$, respectively,
(8-10) the intensity-mean $B$, $V$, and $I$, respectively,
(11-13) the $B$-, $V$- and $I$-amplitude, respectively,
(14) the type of variable, and 
(15) any comments.

The following sections provide the details of the variables in NGC$\,$2808, including 
the redetection of previously unknown variables.

\section{RR Lyrae variables}
All of the RRLs presented in \citet{corwin04} are recovered and found to be 
varying.  Due to the compact nature of NGC$\,$2808, extreme crowding 
and blending issues prevented us from determining magnitudes for some
of the \citet{corwin04} variables.  

The ground-based data presented here gives satisfactory
magnitudes and amplitudes for four (out of ten) stars classified by
\citet{corwin04} as RR0 Lyrae variables, and seven (out of eight) stars 
classified as RR1 (or possible RR1) stars.  Five of these eight possible 
RR1 stars are bona fide first-overtone (FO) pulsators, whereas we could 
not confirm V17 as an RR1 variable.  Also, the two suspect RR1 variables, 
V13 and V24, are found to be too bright to be RRLs belonging to the cluster, 
and are discussed in more detail below, in \S8.  
Time-series {\it HST} 
observations would be useful to resolve the other severely blended variables.
We note that the existing $HST$ observations of
NGC$\,$2808 are too deep for our purposes, as the RRLs are saturated.

\begin{figure}[htb]
\includegraphics[width=1\hsize]{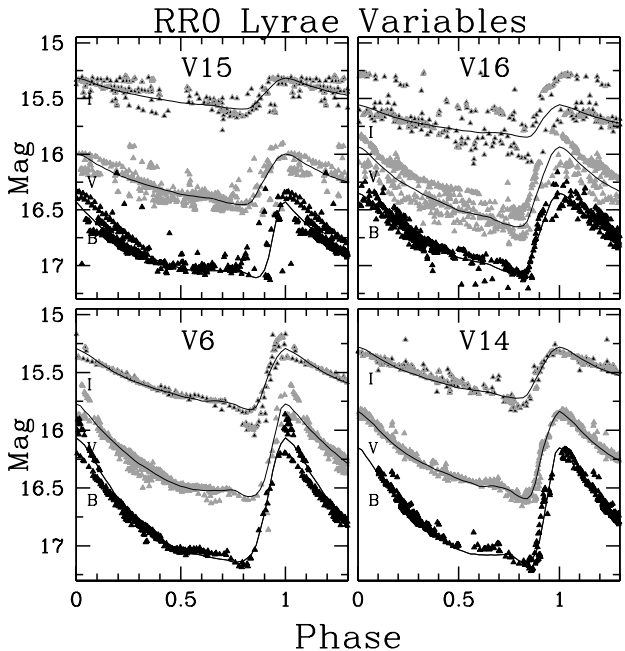}
\caption{Presentation of the phased $BVI$ light curves of fundamental-mode RR Lyrae 
variables in NGC$\,$2808.  
\label{lcrr0}}
\end{figure}

\begin{figure}[htb]
\includegraphics[width=1\hsize]{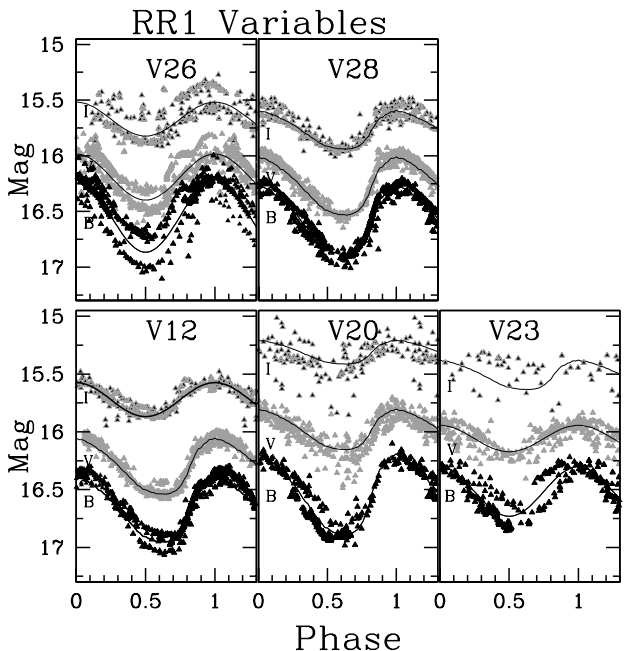}
\caption{Phased $BVI$ light curves of the first-overtone 
RR Lyrae variables in NGC$\,$2808.  
\label{lcrr1}}
\end{figure}

The light curves of the RR0 Lyrae variables are presented in Figure~\ref{lcrr0}, and
the RR1 stars are shown in Figure~\ref{lcrr1}.  The template-fitting routines 
from \citet{layden98} and \citet{layden00} were used to fit the data for the determination
of the pulsation amplitudes.  The mean magnitudes were obtained by averaging 
the results from the largest possible number of cycles.  The difference in the magnitudes
as determined from the template-fitting routines and from averaging over all observing
runs is between 0.001 and 0.03 mag, depending mainly on the scatter about the 
template fit.  Our periods for V17, V19, V21 and V25 are uncertain, as we were unable to find reasonable 
phased light curves for these stars from our data.  Our best period for V17, $\rm P\simeq$0.5 d, suggests
that this star is an RR0 Lyrae star or possibly a double-mode variable (RR01) instead
of an RR1 as suggested by \citet{corwin04}.  The globular clusters M3 and
IC$\,$4499 are known to have RR01 variables with periods falling between
0.46 and 0.50 d \citep{corwin99, clementini04, clement85, walker96},
and V17, V19 and V25 have periods within this general range.

The mean periods for the 11 RR0 Lyrae stars (including V17) and 5 RR1 stars are 
$\langle P \rangle_{RR0}=$0.56$\pm$0.01 d and $\langle P \rangle_{RR1}=$0.30$\pm$0.02 d, 
respectively.  Upon the omission of V17, $\langle P \rangle_{RR0}=$0.56$\pm$0.01 d is found, and also excluding 
V19, V21 and V25, gives $\langle P \rangle_{RR0}=$0.58$\pm$0.01 d.
The ratio of RR1 to total RR Lyrae stars, $N_1/N_{RR}$, is 0.31 (0.38 if V17 is an RR1 Lyrae
star).  As noted in \citet{corwin04}, the mean periods of the RR1 and RR0 Lyrae stars are similar to those
found in the typical Oosterhoff I (OoI) clusters, whereas the $N_1/N_{RR}$ ratio is more in line
with Oosterhoff II (OoII) clusters.   However, Oosterhoff type I systems with similarly high 
RR1-type number fractions are not unprecedented \citep[e.g.,][and references therein]{catelan12}. 


Except for V14, all of the RRLs show evidence of light curve modulation,
likely due to the Blazhko effect \citep{blazhko07}--a poorly
understood effect that appears as a cyclic modulation of shape and amplitude of the light curve
\citep[see e.g.,][for a recent review]{kolenberg11}.
Despite the relatively short time baseline of
the \citet{corwin04} observations (two nights in December 2002, and two nights in 
February 2003), this modulation is seen in many of their light curves as well. 
Assuming there are 11 RR0 Lyrae variables and 5 RR1 variables in the cluster,
a lower limit of 27\% and and an upper limit of
91\% of the RR0 variables exhibit Blazhko behavior, and 100\% of the RR1
exhibit Blazhkocity.  
Period doubling is an effect that would cause light curve scatter predominantly 
at maximum light \citep[e.g.][]{kolenberg10}, and it may be some of these stars (e.g., V6) are
period doubling candidates.  Although our observations span many years, 
the sporadic spacing and sparse consecutive sampling of our observations 
is not sufficient to confirm period doubling or derive Blazhko periods.

Large period-change rates can also introduce scatter in the light curves, and
especially since our images span $\sim$28 years, some of the light curve scatter
may be due to a rapidly changing period.  However, a period change would 
shift the whole light curve in phase space, and not cause the variation of the 
shape and amplitude of the light curve, as is clearly seen in our data.
We separated the light curves into different epochs of photometry and 
attempted to derive periods for the different subsets.  However, these individual subsets 
were insufficient to produce period change rates that could account for the light curve scatter--
the scatter always pointed to an erratic, rather than a secular, period change.
It may also be that some of the RR1 stars are in fact double-mode
pulsators, although our search for secondary periods did not 
reveal any apparent multiple frequencies.

It is worth mentioning that faint neighbor stars may be contaminating the stellar 
PSF profile used to obtain our photometry.  Also, inaccurate entries in the FITS headers of
the archival images may contribute scatter in the light curves.
We stress that we have no specific evidence for contaminated PSF fits, or of 
incorrect header dates, times, or filter identifications in the images employed here.  But we do remain
alert to the fact that PSF fitting in such crowded fields from
instruments with different pixel scales and under different 
observing conditions is challenging, and also the hardware and software that filled in
the metadata for the images were occasionally flawed, especially
in the earlier years.

The Blazhko effect has been shown to occur more often in RR0 variables with 
short periods (P $<$0.55 d, Jurcsik et~al. 2011; see also Preston 1964).
As all the NGC$\,$2808 RR0 Lyrae variables are OoI-type, which on average have shorter
periods than OoII-type variables, a large amount of
Blazhko variability in the RR0 variables is consistent with this result.
As far as the RR1 stars are concerned, a large fraction of Blazhko variables is uncommon, 
but not unprecedented \citep{arellano12}. 

\subsection{The Color-Magnitude Diagram}
\begin{figure}[htb]  
\includegraphics[width=9cm]{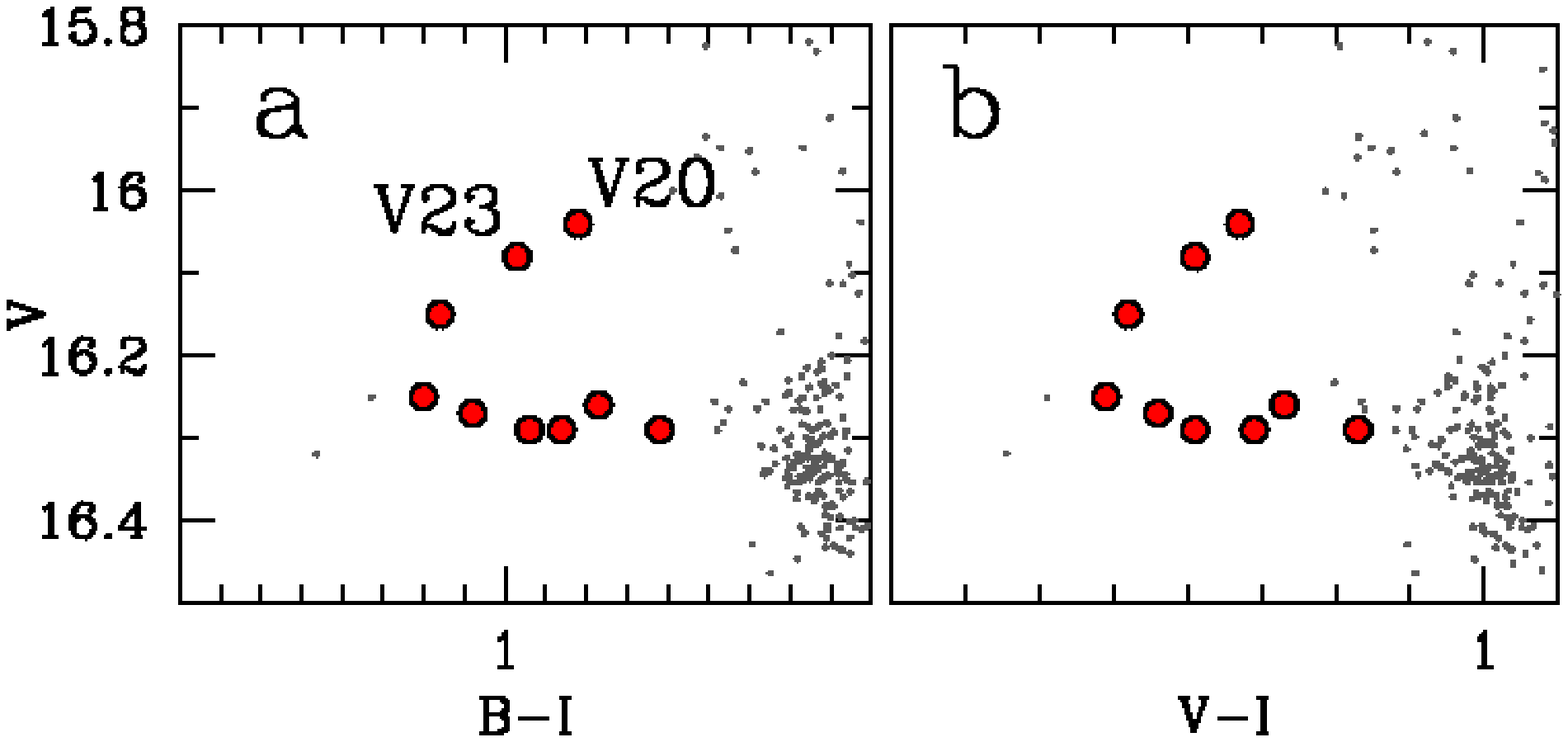}
\includegraphics[height=9cm]{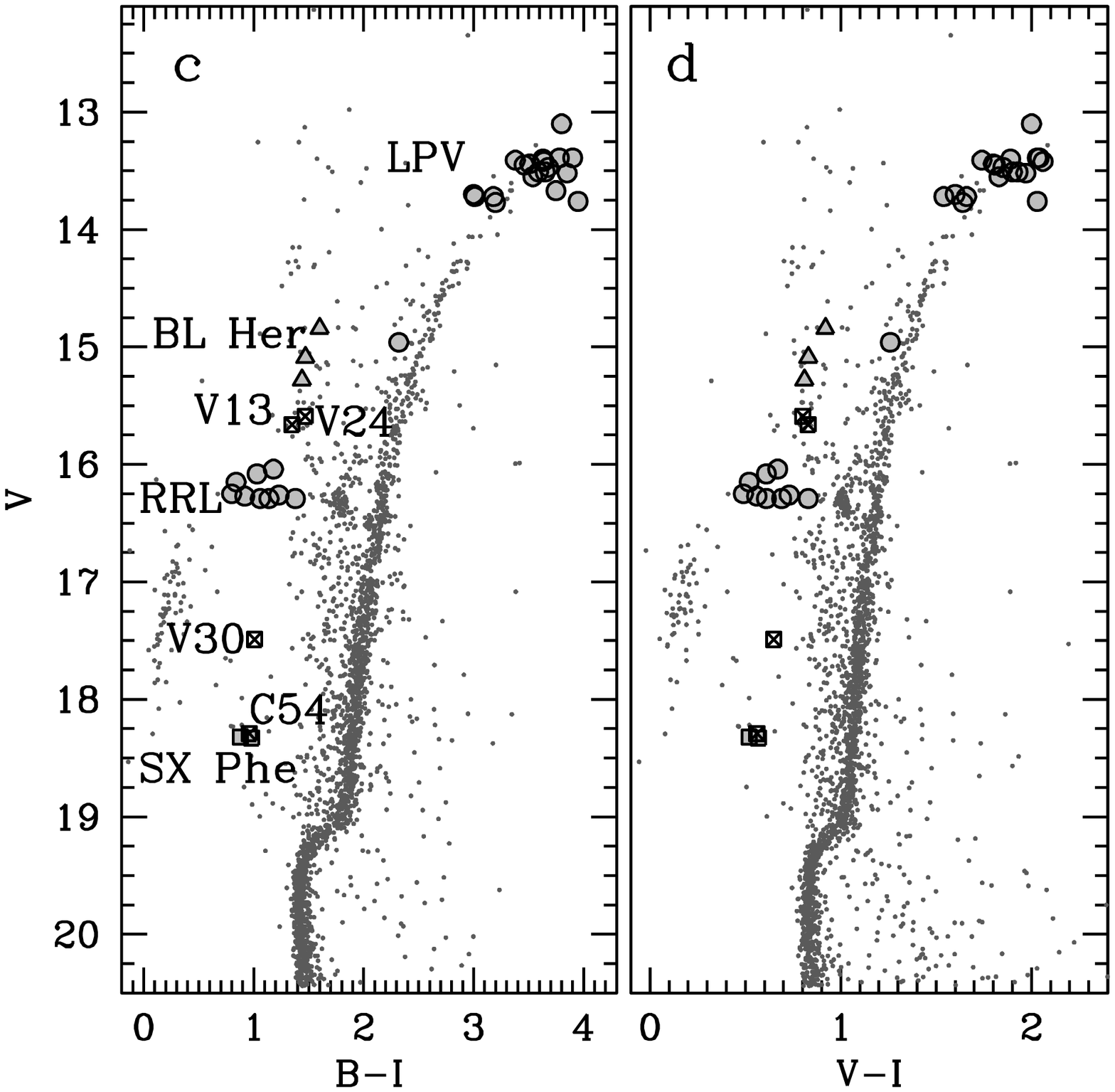}
\caption{NGC$\,$2808 color-magnitude diagram showing the location of the variable stars
studied in this survey.
The bottom two plots [c,d] show those found along the HB are clearly 
RR Lyrae stars, while those found brighter than the horizontal branch are the BL Her stars. 
Those variables along the giant branch are the LPVs.  The SX Phoenicis
stars are the faintest variables in the CMD.  In the top two plots [a,b] a close-up of the HB 
is shown. 
\label{rrcmd}}
\end{figure}

Figure~\ref{rrcmd} shows the positions of the RRLs in the color-magnitude
diagrams for NGC$\,$2808 in $V$ versus (a,c) \hbox{\it B--I\/} and (b,d) \hbox{\it V--I\/}.
Figures~\ref{rrcmd}a and ~\ref{rrcmd}b give an expanded view of the HB.
 
The mean apparent magnitude for the RRLs is $\rm m_{V,RR}$ = 16.21 $\pm$ 0.04 mag,
where the confidence interval is the standard error of the mean.  This is in
excellent agreement with the \citet{walker99} estimate of $\rm m_{V,RR}$ = 16.22, obtained by 
comparing its red HB clump, $V_{\rm RHB}$, 
with that of NGC$\,$1851 and NGC$\,$6362, both of which have a similar $\rm [Fe/H]$ and
assuming $\rm m_{V,RR}$ - $V_{\rm RHB}$=$-$0.08 $\pm$ 0.01.
Adopting \feh=$-$1.14 dex on the \citet{carretta09} metallicity scale, (\feh=$-$1.00 dex 
on the scale of Carretta \& Gratton 1997, \feh=$-$1.05 dex on the Kraft \& Ivans 2003 scale) 
and using the recent recalibration of the RR Lyrae luminosity scale by \citet{catelancortes08}, 
the RR Lyrae variables have an absolute magnitude of
$M_V$=0.70$\pm$0.13 mag.  This leads to an RR Lyrae distance of 
$(m-M)_{V,RRL}$=15.57$\pm$0.13 and using E(\hbox{\it B--V\/})=0.17 mag (see below, \S4.3),
$(m-M)_{0,RRL}$=15.04$\pm$0.13 mag.  
Similarly, using a quadratic relation between RR Lyrae absolute magnitude
and metallicity from Bono, Caputo, \& di Criscienzo (2007), the 
RR Lyrae $M_V$=0.76$\pm$0.08, 
where 0.08 is a reasonable error in the RR Lyrae absolute magnitude-\feh zero-point
calibration.  This leads to an $(m-M)_{V,RRL}$=15.51$\pm$0.09 and
$(m-M)_{0,RRL}$=14.98$\pm$0.09 mag. 

\begin{figure}[h!]
\begin{center}
\subfigure[V23]{  
\includegraphics[height=5cm]{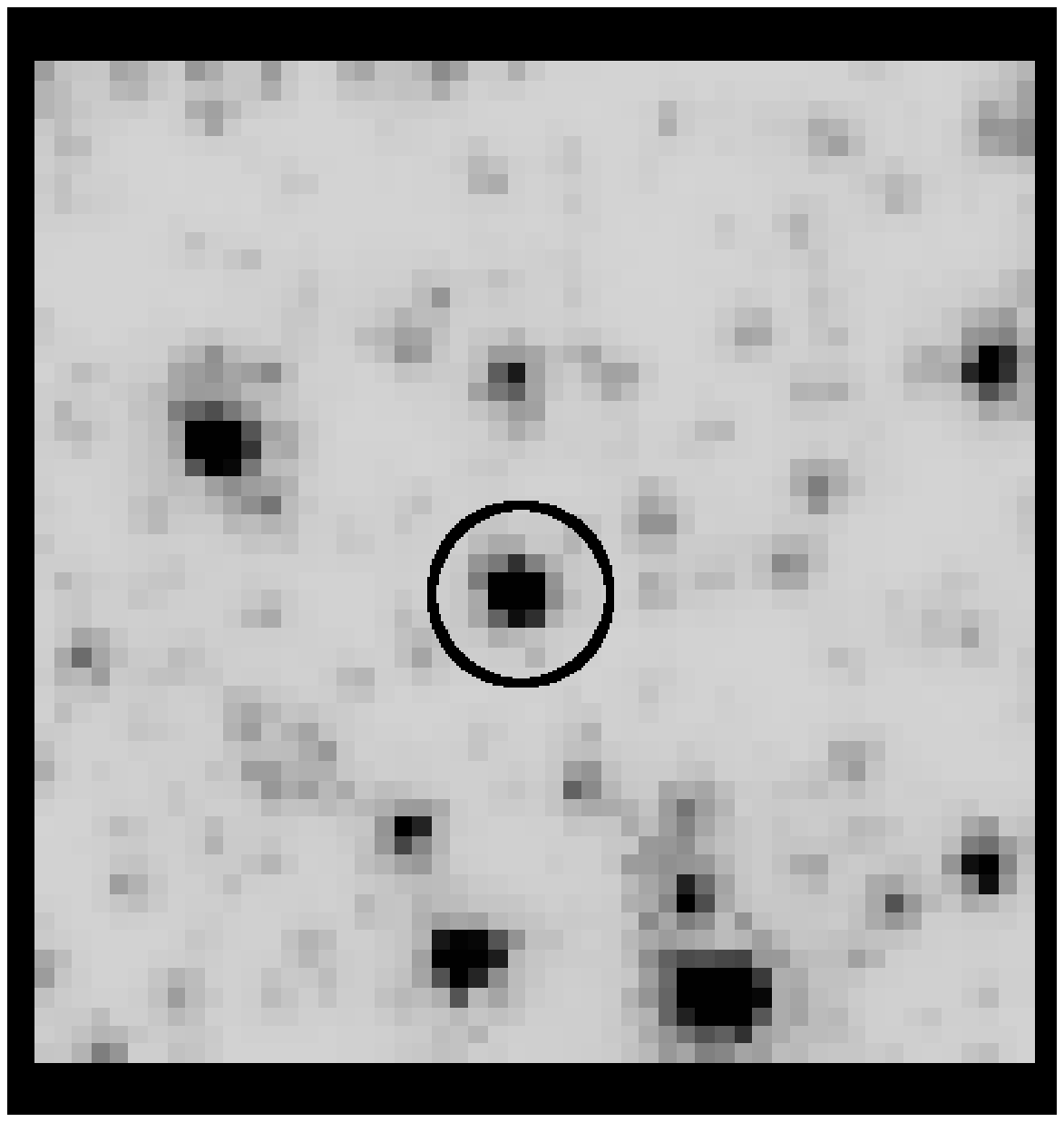}}
\subfigure[V20]{
\includegraphics[height=5cm]{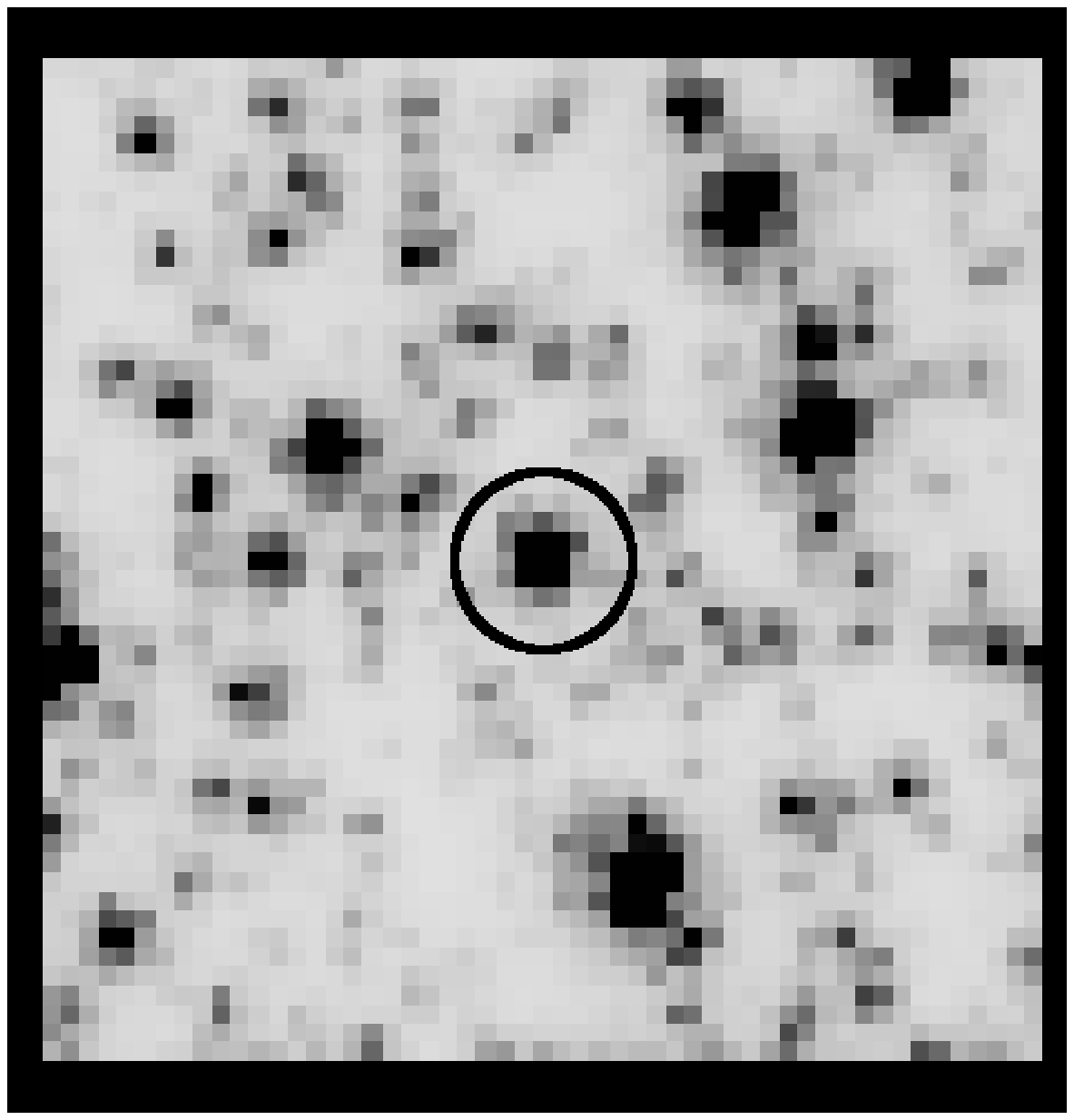}}
\caption {\label{v20v23} Stack of all available $HST$ images of NGC$\,$2808 centered on (a) V23 and
(b) V20, the two brightest RR Lyrae variables in our sample.  Each image is 4 arcseconds on 
a side, where North is up and East is left.
}
\end{center}
\end{figure}

As seen in Figure~\ref{rrcmd}, two stars are found at magnitudes brighter than the majority of 
the RRL stars. These are the RR1-type variables V20 and V23, and they are
$\sim$0.25 mag brighter than the rest of the variables.  We do not believe these stars
are blended, although they both lie fairly close to the center of the cluster.  Figure~\ref{v20v23} shows
thumbnails of a stacked $HST$ image centered on these stars.  There is no bright neighboring star 
within $\sim$1.5 arc seconds of V23, and V20 is about 1.0 arcseconds away from three relatively bright stars.
Although this could contribute to slight blending, the contamination is at the level of a few percent---not much 
more than the uncertainty in the photometry.  Because these stars have colors that  
overlap with the bluest part of the RR0 Lyrae region in the CMD, they may reside
in the {\it either-or\/} region of the IS, where the transition between RR1 and RR0 takes place
and where both first-overtone and fundamental-mode RRLs can 
be found \citep[e.g.,][]{vanalbada73, caputo78, stellingwerf75, bono95a}. 
It may be that these stars have evolved further from the zero-age HB and are thus more luminous. 
Their periods, however, are not abnormally long for RR1 variables.  We also do not see any evidence
for large secular period change rates, but note that finding period change rates of Blazhko stars
is an especially challenging task requiring large amounts of data \citep{kunder11}.  
\citet{sandage90} show that metal-rich clusters tend to exhibit a larger spread in RR Lyrae 
absolute magnitude ($\sim$0.5 mag for $\rm [Fe/H]$=-1.01), and the spread in $V$ we find for the
RR Lyrae variables is consistent with this estimate.

\begin{figure}[htb]  
\includegraphics[height=9cm]{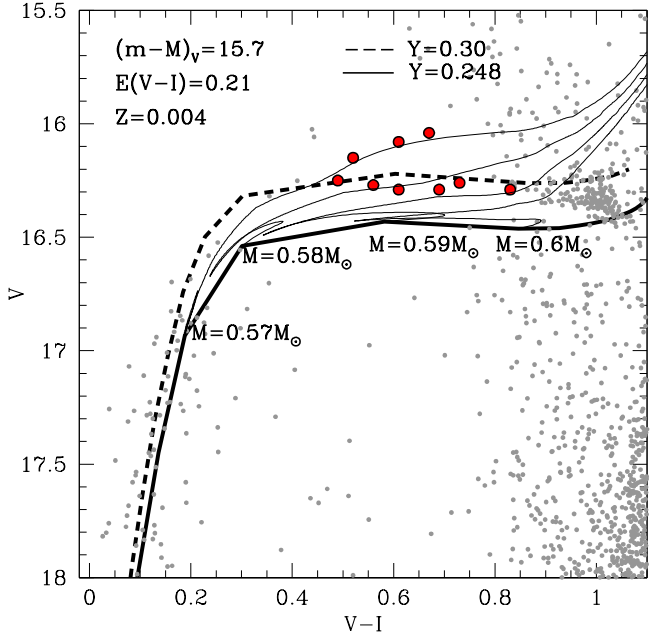}
\caption{Optical CMD of the cluster HB compared to theoretical ZAHBs for $Y$=0.248 (heavy solid) 
and $Y$=0.30 (dashed).  BaSTI HB evolutionary models are also shown for M=0.58$M_\sun$ and 
M=0.60$M_\sun$ HB stars with $Y$=0.248.
\label{RRiso}}
\end{figure}

Figure~\ref{RRiso} shows the $V$,$\hbox{\it (V--I)\/}$ CMD with the blue HB, the red HB and
the RRL IS.
Two Zero Age Horizontal Branches (ZAHBs) from the BaSTI $\alpha$-enhanced HB 
tracks \citep{pietrinferni04, pietrinferni06} with $\rm [Fe/H]$=$-$1.01 dex are over-plotted.
One has a helium abundance of $Y$=0.248 and the other has $Y$=0.30.
An apparent distance modulus of $\rm (m - M)_V$=15.7~mag 
matches the observed mean magnitude of the red HB for the normal He track.  
This is $\sim$0.13 mag fainter than the distance modulus 
obtained from the RR Lyrae variables.
One always expects the evolutionary mean level of the 
RR Lyrae stars to be different from the ZAHB, as shown 
by e.g., \citet{catelan92}.  For the metallicity of NGC$\,$2808,
the difference between the ZAHB and the average RR Lyrae magnitude is
predicted to range from $\sim$0.18 mag \citep{catelan92} to $\sim$0.1 \citep{gallart05}, 
consistent with what is found here.
Therefore, a ZAHB with a primordial He abundance fits the observations
well, suggesting that the majority of the RRL instability strip is not helium enhanced.
We also show the BaSTI evolutionary tracks for 0.57$M_\sun$ and 0.60$M_\sun$ HB stars. 
The BaSTI tracks also indicate that most of the NGC$\,$2808 RR Lyrae stars have a rather small mass
range of 0.57 -- 0.60$M_\sun$.

\subsection{Period-Amplitude Diagram}
\begin{figure}[htb]  
\includegraphics[width=1\hsize]{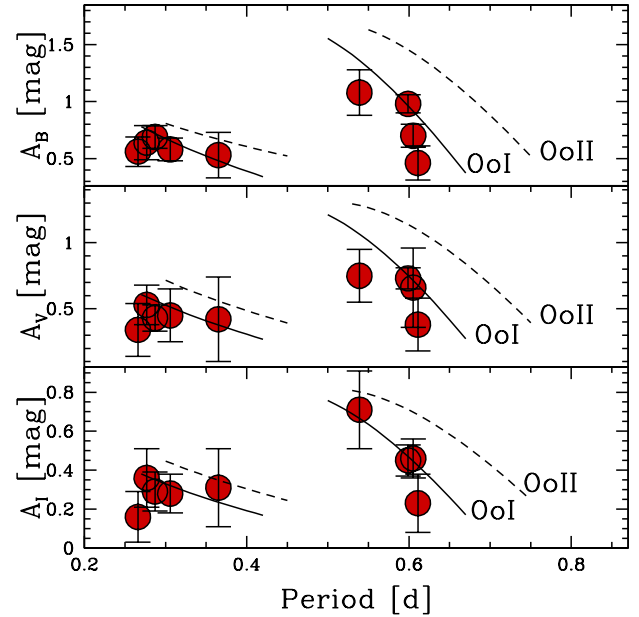}
\caption{Position of RR Lyrae stars on the period-amplitude diagram for $V$ (bottom), $B$ (middle)
and $I$ (top).  Solid lines are the typical lines for OoI clusters and dashed lines for OoII clusters, 
according to \citet{cacciari05} and \citet{zorotovic10}.  Because all of these stars exhibit
the Blazhko effect, the error-bars designate the change in amplitude
from the average light curve amplitude to the maximum Blazhko amplitude.
\label{PA}}
\end{figure}

Figure~\ref{PA} shows the period-amplitude (PA) diagram for the RR Lyrae stars.  
Also shown in the $B$- and $V$-amplitude plane are typical lines for OoI and OoII 
clusters, where the fundamental-mode PA
relation is derived from the M3 RRL  \citep{cacciari05} and the first-overtone
PA relation is derived from the NGC$\,$5286 RRL \citep{zorotovic10}.
The Oosterhoff loci of the RRLs in the $I$-amplitude plane is determined
using $A_V$ = 1.6 $\times$ $A_I$, as determined from Table 6 of \citet{liu90}.
These relations then become:
\begin{equation}
A_I^{RR0} = -1.64-13.78 \log P - 19.30 \log P^2
\end{equation}
\begin{equation}
A_I^{RR1} = -0.25 - 1.10 \log P
\end{equation}
for fundamental-mode and first-overtone RR Lyrae stars in OoI clusters and
\begin{equation}
\begin{multlined}
A_I^{RR0} = -1.64-13.78 (\log P - 0.04) \\ 
- 19.30 (\log P - 0.04)^2
\end{multlined}
\end{equation}
\begin{equation}
A_I^{RR1} = -0.15 - 1.15 \log P
\end{equation}
for RR Lyrae stars in OoII clusters.
The period-$I$-amplitude relation derived for the OoII stars in M53
by \citet{arellano12} has a shallower slope 
that extends to longer periods, so the NGC$\,$2808 RRLs fall even 
farther from the \citet{arellano12} OoII relation than to the one shown in Figure~\ref{PA}. 

The variables all occupy the area of the PA diagram where the M3 RRL as well
as where other Oosterhoff I RRLs have been shown to reside \citep[e.g.,][]{clement99}.  
The position of a star in this diagram can be affected by the presence of 
the Blazhko effect \citep[e.g.,][]{cacciari05}, and as discussed above, almost 
all the RR Lyrae variables presented here show signs of this phenomenon.  Because the light 
curves presented here have a large number of observations, it is straightforward 
to determine the amplitudes using the average light curves of the Blazhko stars.
Nevertheless a visual determination of the change in amplitude in each Blazhko RR0 is 
obtained and shown as a symmetric error-bar in the figure.  The change from the 
average light curve amplitude to the maximum Blazhko amplitude ranges 
from $\sim$0.1 - 0.4 mag.  Even when taking the amplitude variation
due to the Blazhko cycle into account, the variables are all OoI-type.
We note that our assumption of a symmetric Blazhko amplitude change may 
overestimate the lower amplitude value, as exhibited for example by the 
shortest-period RR Lyrae variable V23.  The light curve of V23 does not seem to 
show amplitudes that go down to zero at some points; its error-bar reaching to 
almost zero is an artifact of assuming symmetric error bars.  Such an overestimate 
would not change our conclusion that all the variables are OoI-type.
\subsection{RR Lyrae Reddenings}
The interstellar extinction toward NGC$\,$2808 is relatively large 
\citep[E(\hbox{\it B--V\/}) = 0.19 mag; 2010 edition of][]{harris96} and differential reddening is 
also thought to affect the stellar colors \citep[$\rm \Delta E(\hbox{\it B--V\/})$=0.02][]{piotto07}.  Recent 
$\rm E(\hbox{\it B--V\/})$ values used for this cluster vary from $\rm E(\hbox{\it B--V\/})$ = 0.13 \citep{castellani06} to 
$\rm E(\hbox{\it B--V\/})$ = 0.23; \citep{piotto02}.  Here the minimum light 
colors of the RR0 Lyrae variables are used to investigate the reddening of NGC$\,$2808.

Fundamental-mode RRLs are known for their dramatic light variations, one of
the reasons these stars are a particularly fun target to observe.  At minimum light, however, 
RR0 Lyrae stars exhibit a much narrower temperature range \citep{preston59, lub77, fernley96}.
This property has been used as a tool for measuring the interstellar reddening toward
RRLs \citep{sturch66, blanco92, mateo95, kunder10}.
\citet{guldenschuh05} showed that at minimum light (phase between 0.5 and 0.8) the color of
RR Lyrae stars is $\hbox{\it (V--I)\/}_{0,min}$ = 0.58 $\pm$ 0.02, and that this property is largely 
independent of period, amplitude and metallicity.  

The observed \hbox{\it V--I\/} color at minimum light was calculated from both the best-fit 
light-curve templates as well as the observed data.  The values agreed to within 
0.01 mag, and are listed in Table~\ref{evmi}.  The number 
of data points in our $V$ and $I$ light curves that fall between a phase of 0.5 -- 0.8 is 
between 35 and 200.  

\begin{table}
\begin{scriptsize}
\centering
\caption{Reddening Determinations from RR Lyrae variables in NGC$\,$2808}
\label{evmi}
\begin{tabular}{p{0.3in}p{0.5in}p{0.55in}} \\ \hline
Name & ($\hbox{\it V--I\/})_{min}$ & $\rm E(\hbox{\it V--I\/})$ \\ 
\hline
V6 & 0.76 & 0.18 \\
V14 & 0.81 & 0.23 \\
V15 & 0.84 & 0.26 \\
V16 & 0.77 & 0.19 \\
\hline
\end{tabular}
\end{scriptsize}
\end{table}
The average $\rm E(\hbox{\it V--I\/})$ is 0.215 $\pm$ 0.02, or assuming a standard reddening law
\citep[e.g.,][]{cardelli92}, 
$\rm E(\hbox{\it B--V\/})$ is 0.17 $\pm$ 0.02. This is slightly smaller, but still consistent with
the reddening found by \citet{bedin00} and \citet{walker99}.  It is also smaller than the 
\citet{schlegel98} estimate of E$(\hbox{B--V\/})$= 0.23, but the latter reduces to 0.18~-- in 
excellent agreement with our results~-- after applying the correction suggested by 
\citet{bonifacio00}.  Our $\rm E(\hbox{\it B--V\/})$ value agrees 
remarkably well with the values adopted by \citet{dantona04}, \citet{dalessandro11} 
and \citet{milone12} when fitting isochrones to NGC$\,$2808.

\citet{walker98} showed that the blue edge of the instability
strip in GCs appears to have a constant \hbox{\it B--V\/} over a wide range of metallicity,
with $(\hbox{\it B--V\/})_{0,FBE}$ = 0.18 $\pm$ 0.01.  \citet{mackey03} find 
$(\hbox{\it V--I\/})_{0,FBE}$ = 0.28 $\pm$ 0.02, based on two Fornax GCs.  Since 
then \citet{sandage06} found that a small dependence with $\rm [Fe/H]$ exists at the 
blue fundamental edge, with $(\hbox{\it B--V\/})_{0,FBE}$ varying
parabolically by 0.02 mag between $\rm [Fe/H]$ = $-$0.08 and $-$2.3 dex.  
The bluest RRL in our admittedly small sample has \hbox{\it B--V\/} = 0.31 mag 
and \hbox{\it V--I\/} = 0.49 mag.  
This suggests an upper limit of $\rm E(\hbox{\it B--V\/})$ = 0.13 $\pm$ 0.02 if using the value of the 
FBE from \citet{walker99} and $\rm E(\hbox{\it V-I\/})$ = 0.21 $\pm$ 0.02 or 
$\rm E(\hbox{\it B--V\/})$ = 0.17 $\pm$ 0.02 if using the value of the FBE from \citet{mackey03}.
The color excess values derived from both the blue edge of the IS and
from the RR Lyrae minimum light colors are smaller than the generally
adopted reddening value for NGC$\,$2808 of $\rm E(\hbox{\it B--V\/})$$\sim$0.19.  
To conclude this paragraph it is worth mentioning that the mean color of an 
RR Lyrae star can be different from the color of the ``equivalent static star" (ESS), 
and the ESS color is the one that is used in such applications. Close to the blue edge of 
the IS, the difference between different definitions of mean colors and the color of the ESS 
is of order 0.01 mag -- see Table 4 in Bono et~al. (1995b).
This would not change our conclusions that a slightly smaller
value for $\rm E(\hbox{\it B--V\/})$ is in agreement with the results from the RRL minimum light colors.
\section{Long-Period Variables}
LPVs are radially pulsating stars on the giant branch.
Especially because their pulsation is one of the several linked phenomena
that control the endpoint of stellar evolution, LPV investigations provide important
constraints on models of stellar evolution as well as the injection of mass into the 
interstellar medium \citep[e.g.,][]{arndt97, mennessier01, mcdonald10, lebzelter11}.  
LPVs generally show periodic variations in brightness with periods of
$\sim$30 up to a few thousands of days and amplitudes ranging from
several tenths to approximately ten magnitudes in the visual.
Recently \citet{lebzelter11} identified twenty LPVs in NGC$\,$2808 and claimed that a high 
helium abundance of $Y$$\sim$0.4 is required to explain the periods of several of the 
LPVs.  Irregularities in the light change were noted on top of the periodic variations. 

\begin{figure}[htb]
\includegraphics[width=1\hsize]{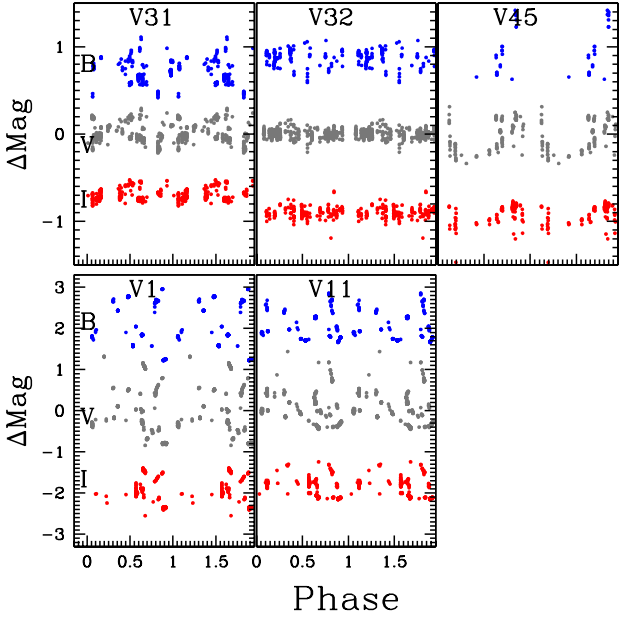}
\caption{Presentation of the phased light curves of LPVs in NGC$\,$2808.  The
$B$ (top), $V$ (middle) and $I$ (bottom) light curves are shown in the same panel.  The
$B$ and $I$ light curves are offset by an arbitrary amount for clarity.
\label{lpv1}}
\figurenum{1}
\end{figure}

The {\it BVI\/} light curves for these stars, phased with the periods obtained by \citet{lebzelter11}, 
are shown in Figure~\ref{lpv1}.  Their mean magnitudes are listed in Table~\ref{lcpars}.  
Unfortunately our  time coverage is not suitable for a definitive 
investigation of the LPVs, and we do not derive periods or discuss these stars in 
detail here.  However, as Figure~\ref{varindex} shows, we do note that the W/S variability index 
\citep{welch93} tends to confirm that the LPVs are indeed variable.  Some 
of the periods reported by \citet{lebzelter11} failed to phase our light curves properly, but 
it is difficult to ascertain whether the periods are wrong or whether the scatter is merely caused by 
irregularities in the light curves.  Especially since the three LPVs with the longest periods,
as found by  \citet{lebzelter11}, were used to infer a $Y$$\sim$0.4 to explain their LPV observations,
more epochs would be useful.  
\begin{figure}[htb]  
\includegraphics[width=1\hsize]{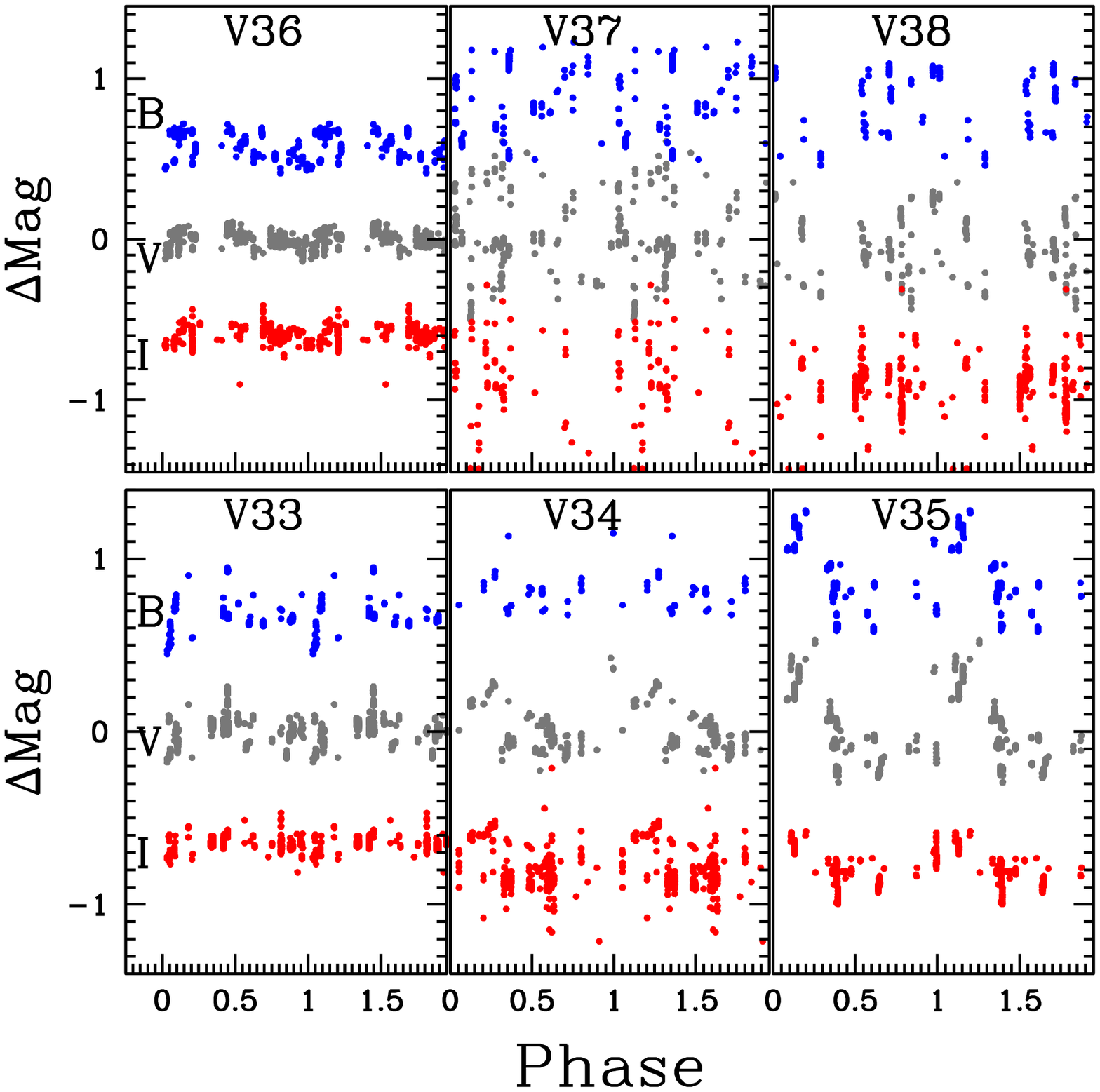}
\caption{LPVs in NGC$\,$2808 continued
\label{lpv2}}
\figurenum{1}
\end{figure}
\begin{figure}[htb]  
\includegraphics[width=1\hsize]{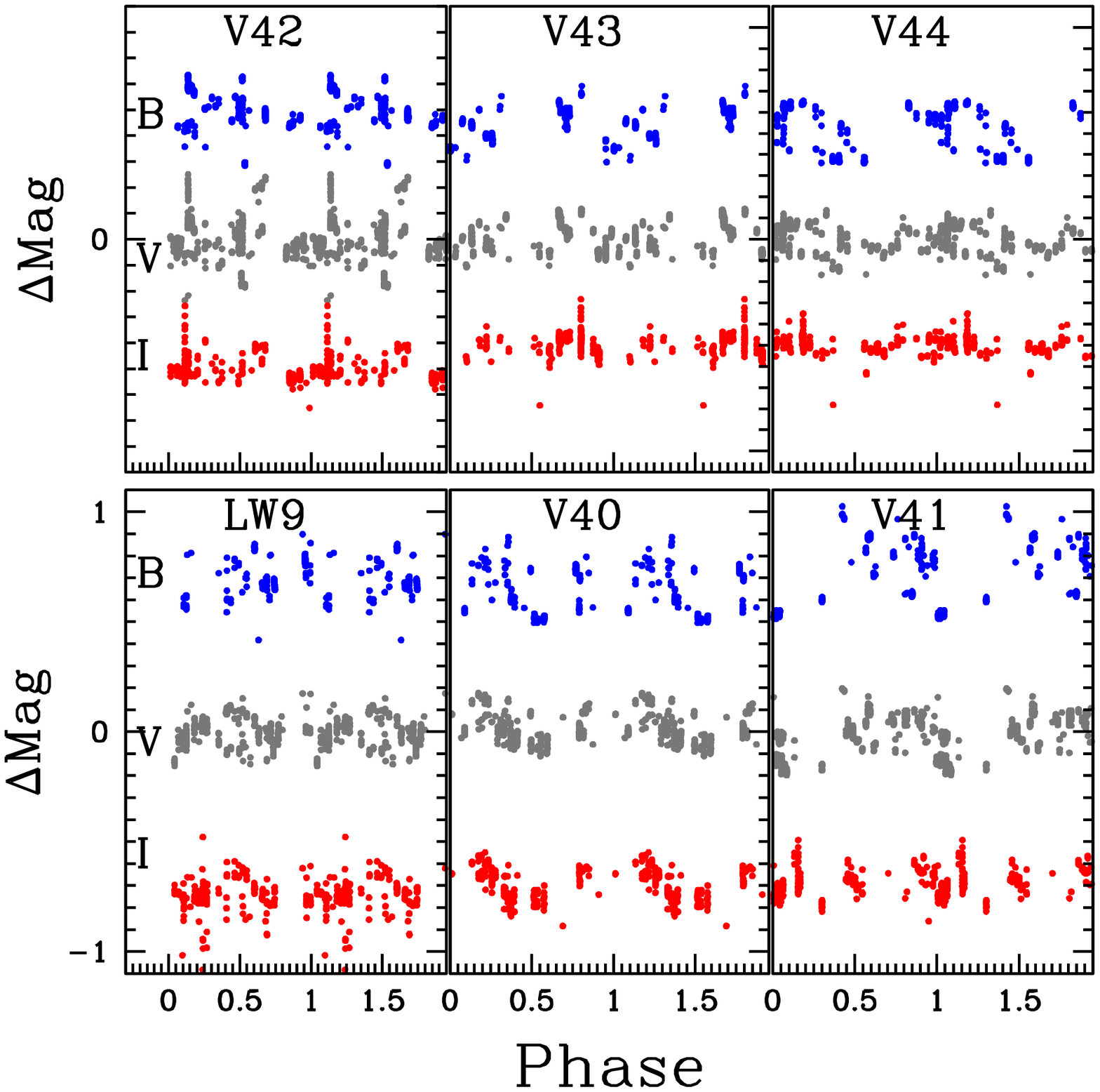}
\caption{LPVs in NGC$\,$2808 continued
\label{lpv3}}
\end{figure}

\begin{figure}[htb]  
\includegraphics[width=1\hsize]{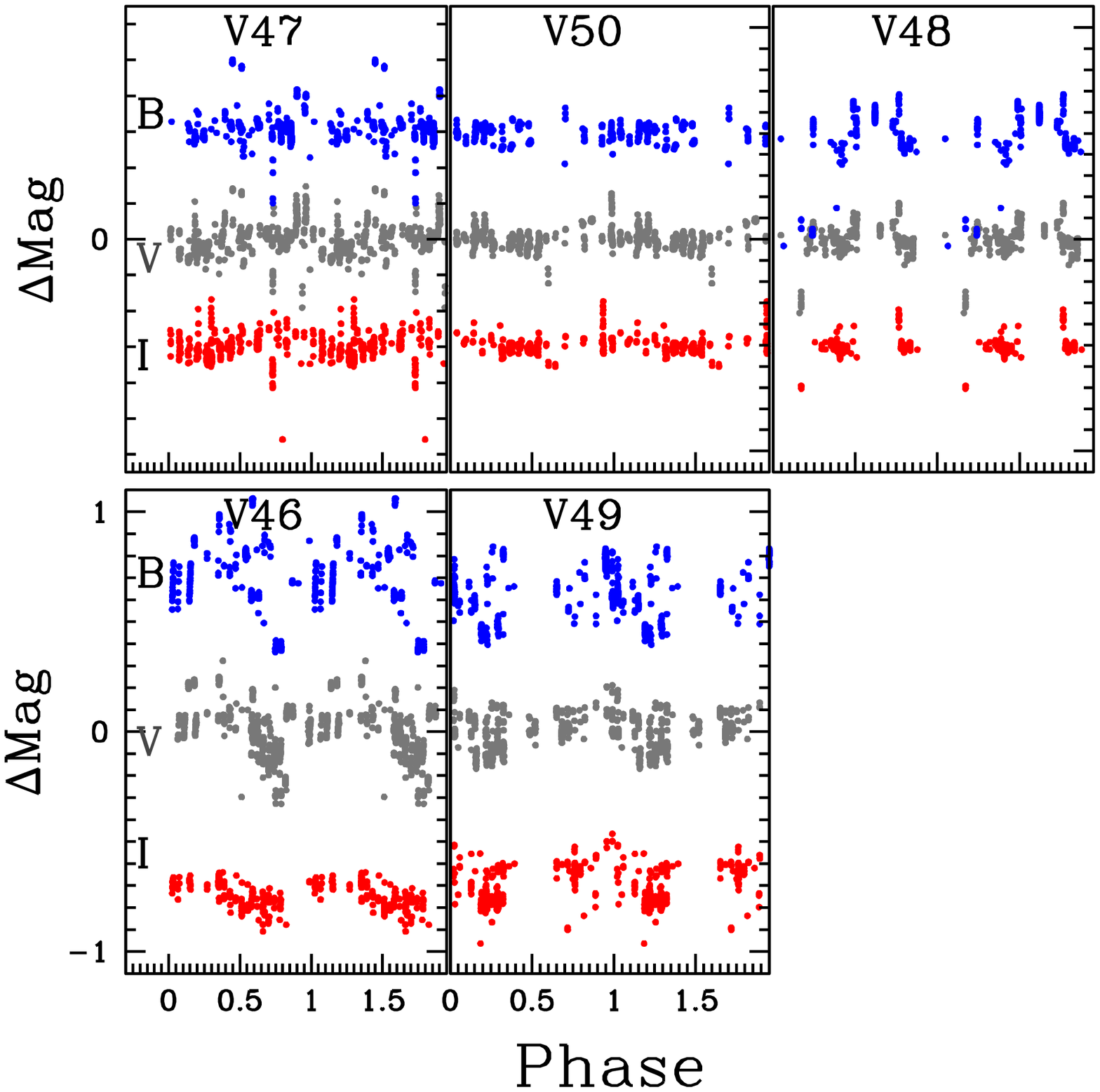}
\caption{The Welch-Stetson variability index for the LPVs, the non-variable stars, and the BL Her and RRL stars
is plotted against the $V$ magnitude.  The non-variable star, V3, also has a relatively large W/S index.
\label{varindex}}
\end{figure}

\section{SX Phe Stars}
SX Phoenicis stars (SX Phe) are of particular interest in Galactic GCs, because 
the region of the instability strip which they occupy is coincident with the 
blue straggler region \citep[e.g.,][]{jeon04}.  This means that they have an unusual 
life history which is not well understood as it cannot be explained in terms of the standard 
single star evolution scenario.  One method that may assist in the as-yet-unsolved issues of 
blue straggler star (BSS) formation is an analysis of their pulsation frequencies, which
can provide information on the interiors of SX Phe stars \citep[e.g.][]{bruntt01, olech05,
cohen12}. 

\begin{figure}[htb]  
\includegraphics[width=1\hsize]{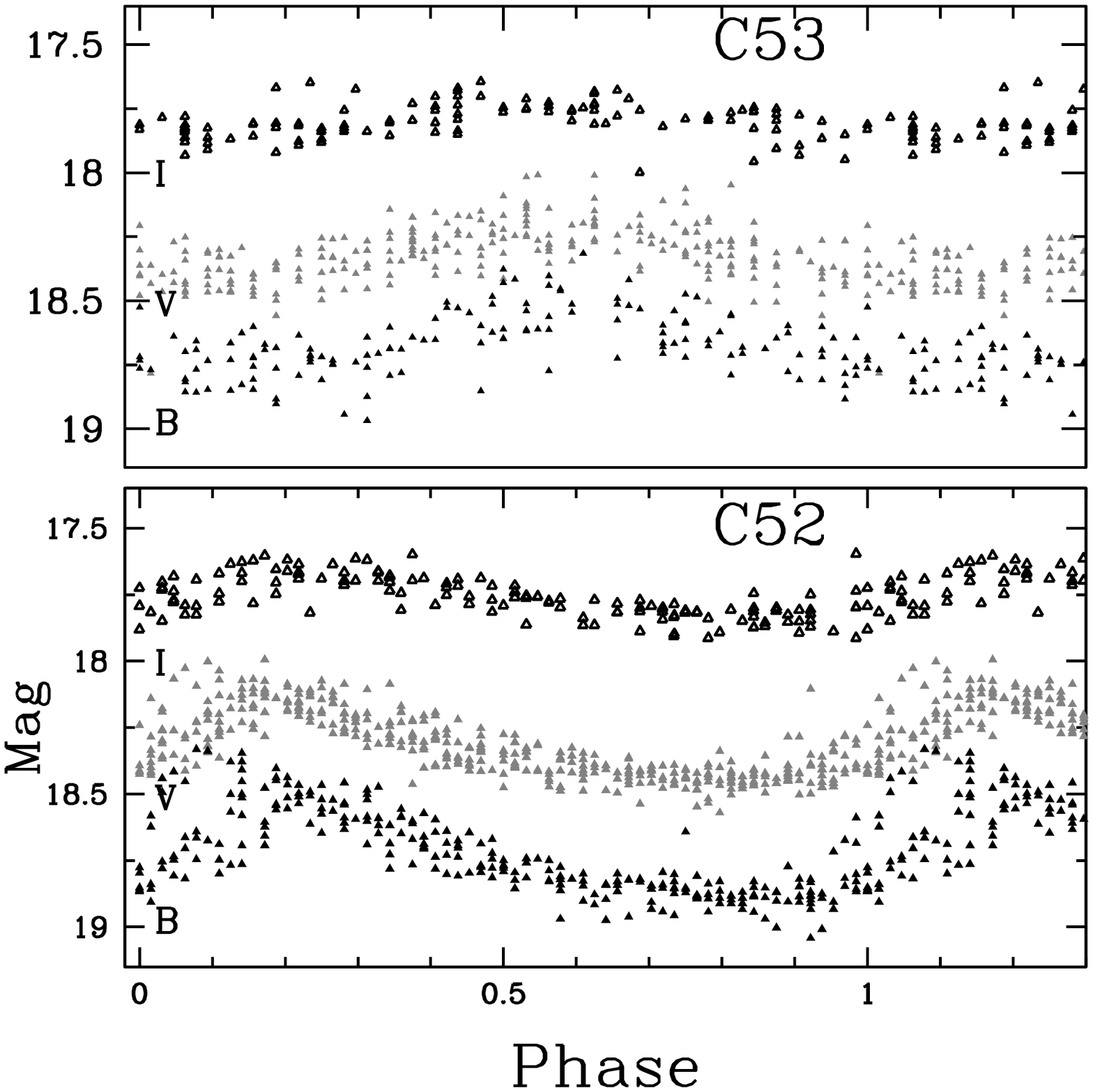}
\caption{Phased $BVI$ light curves of the SX Phe stars identified in NGC$\,$2808.  The primary
period is used to phase the light curve of C53.
\label{sxphe}}
\end{figure}

\begin{figure}[htb]  
\includegraphics[width=1\hsize]{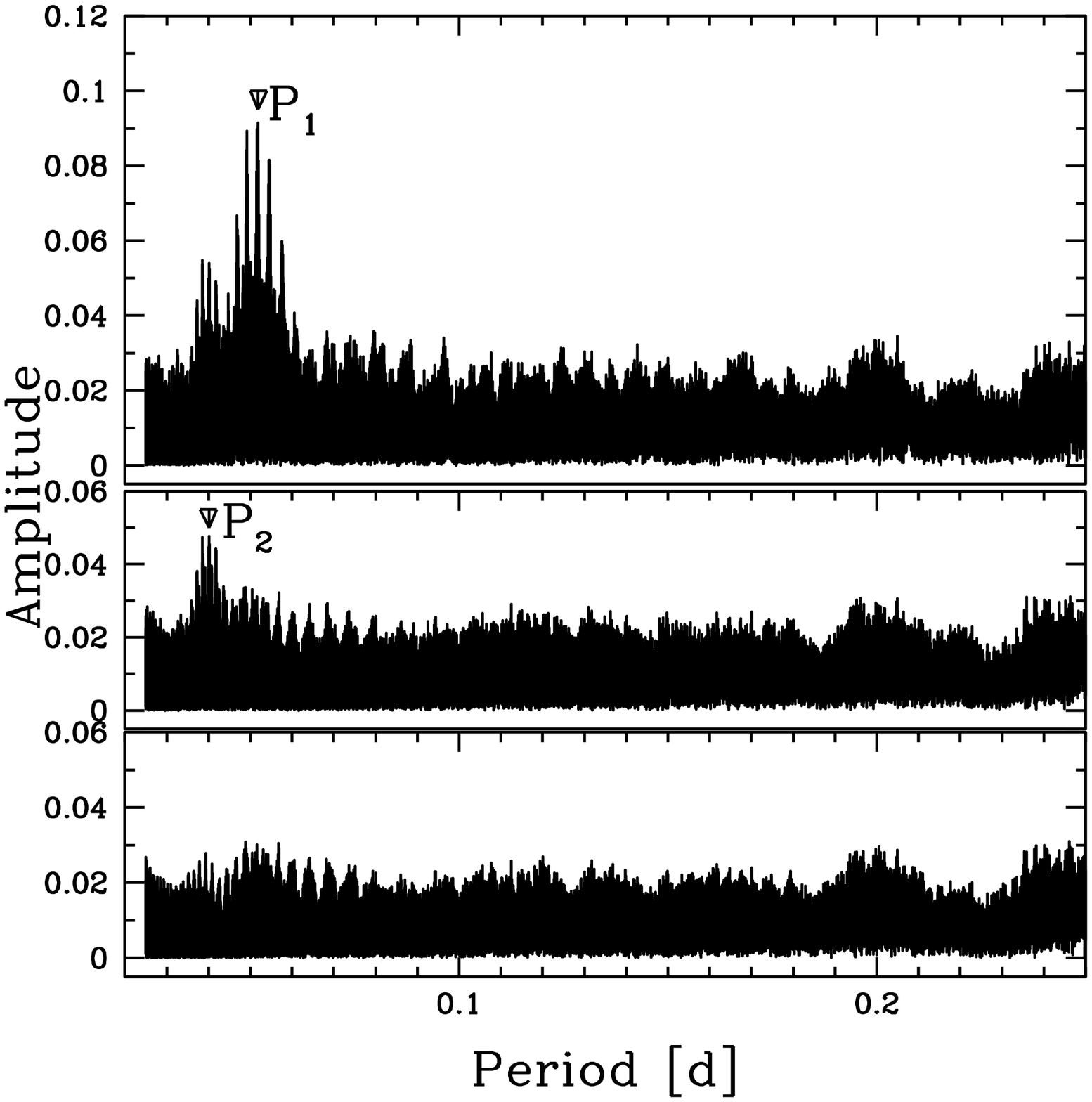}
\caption{Fourier spectra of C53: {\it(top)} for original $V$-filter observations, (middle) after
pre-whitening with period $P_1$ = 0.05181966 d, (bottom) after removing terms with
periods $P_1$ and $P_2$ = 0.04011174 d.  The ordinate scales in all panels are the same.
\label{c53power}}
\end{figure}

We have identified the first candidate SX Phe stars in NGC$\,$2808, and their light curves are
shown in Figure~\ref{sxphe}.   Many SX Phe stars exhibit multiple periods and
it is likely that C53 has multiple frequencies as well.  The amplitude spectra of
C53 for original data and after pre-whitening with subsequent frequencies is shown in
Figure~\ref{c53power}.  The ratio of the periods is equal to $P_1/P_0$ = 0.77, in good
agreement with observed double mode pulsators that oscillate in the fundamental mode
and the first overtone \citep[e.g.,][]{alcock00a, pigulski06, mcnamara07, garg10}.
The theoretical period-ratio of the first-overtone to the fundamental
radial mode for SX Phe pulsators with Z=0.001 ($\rm [M/H]$=$-$1.27) falls in
the range from 0.77 to 0.79 \citep{santolamazza01}, with $P_1/P_0$ increasing with increasing
$\rm [M/H]$.  As NGC$\,$2808 has Z$\sim$0.002, the period ratio obtained here is well in line
with the theoretical period-ratio for double-mode SX Phe stars as well.

\begin{figure}[htb]  
\includegraphics[width=1\hsize]{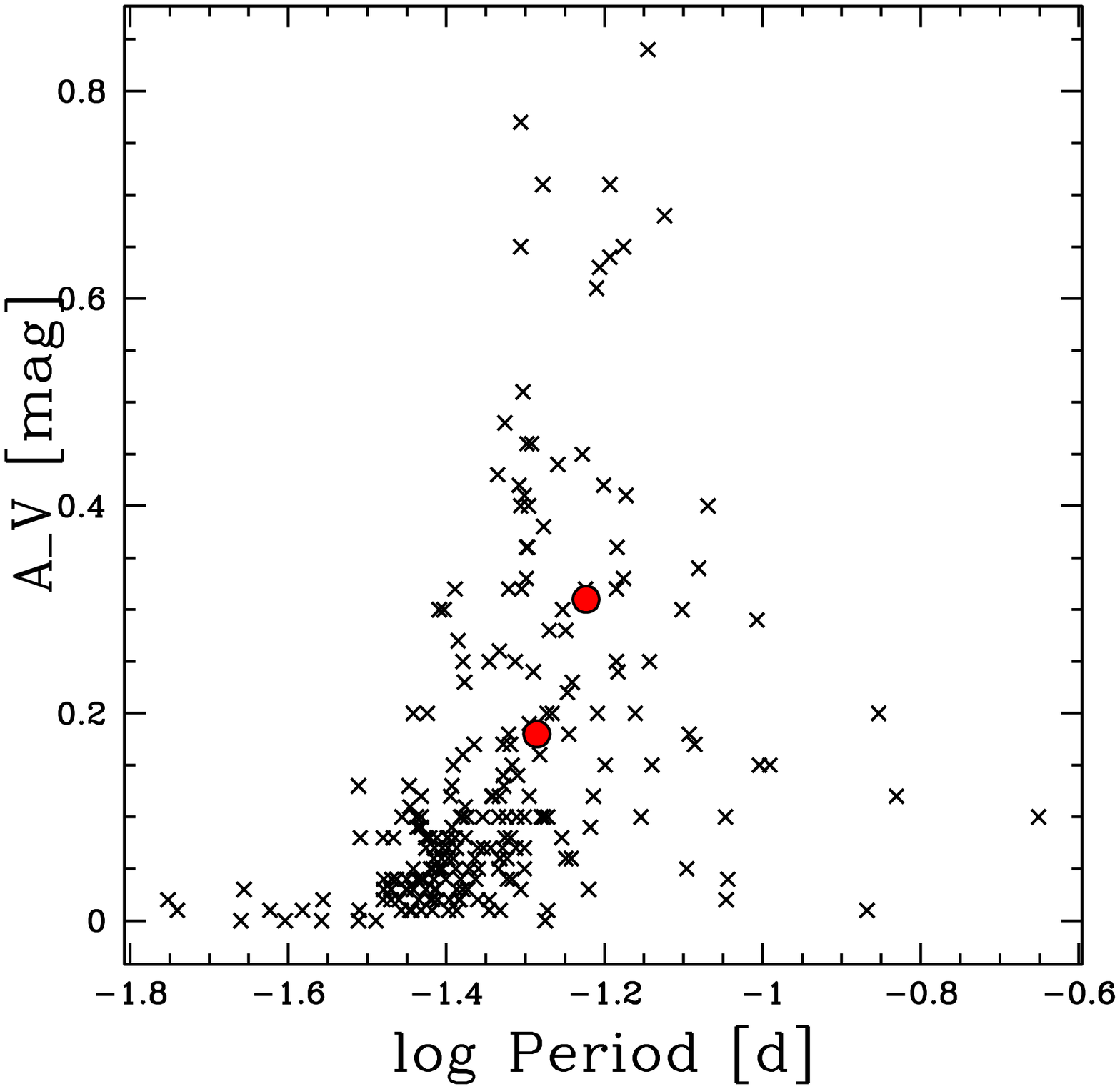}
\caption{The NGC$\,$2808 SX Phe stars periods and $V$-amplitudes (filled circles)
as compared to the periods and amplitudes of SX Phe stars from 27 different Galactic GCs (crosses) 
compiled in \citet{cohen12}.
\label{sxall}}
\end{figure}

Figure~\ref{sxall} indicates that the periods and $V$-amplitudes of the SX Phe 
stars in NGC$\,$2808 are consistent with those in other Galactic GCs.
Figure~\ref{rrcmd} shows the positions of the two SX Phe stars in the color-magnitude diagram.
As expected, the SX Phoenicis stars are located in the blue straggler region, 
brighter and bluer than the main-sequence turnoff point. 

Using 153 $\delta$ Scuti and SX Phe stars in $\omega$ Cen, M55, Carina, and Fornax, \citet{poretti08}
derived an SX Phe period--luminosity relation

\begin{equation}
M_V = -1.83 - 3.65 \log P,
\end{equation}
\noindent
with a standard deviation of $\sigma$ = 0.18 mag.  They noted that scatter in this 
relation is likely caused by the existence of different pulsation modes in many SX Phe stars---which are not always 
straightforward to identify---and the presence of subluminous SX~Phe stars.  Using this relation and the
SX~Phe stars found here, the average distance modulus to NGC$\,$2808 is 
$\rm (m-M)_{V,SXPhe}$ = 15.58 $\pm$ 0.12 mag, where the confidence interval is the standard error 
of the mean.  \citet{cohen12} derived a period--luminosity relation from double-mode 
SX Phe stars in 27 Galactic GCs, and using their relation with our double-mode SX Phe star, C53,
a distance modulus of $\rm (m-M)_{V,SXPhe}$ = 15.60 $\pm$ 0.10 mag is found, where the error is the residual in the
\citet{cohen12} period--luminosity fit.
The two SX Phe distance estimates are in excellent agreement with each other and are also similar to the distance
found using the RR Lyrae variables. 

\clearpage

\clearpage
\section{Population II Cepheids}
\begin{figure}[htb]  
\includegraphics[width=1\hsize]{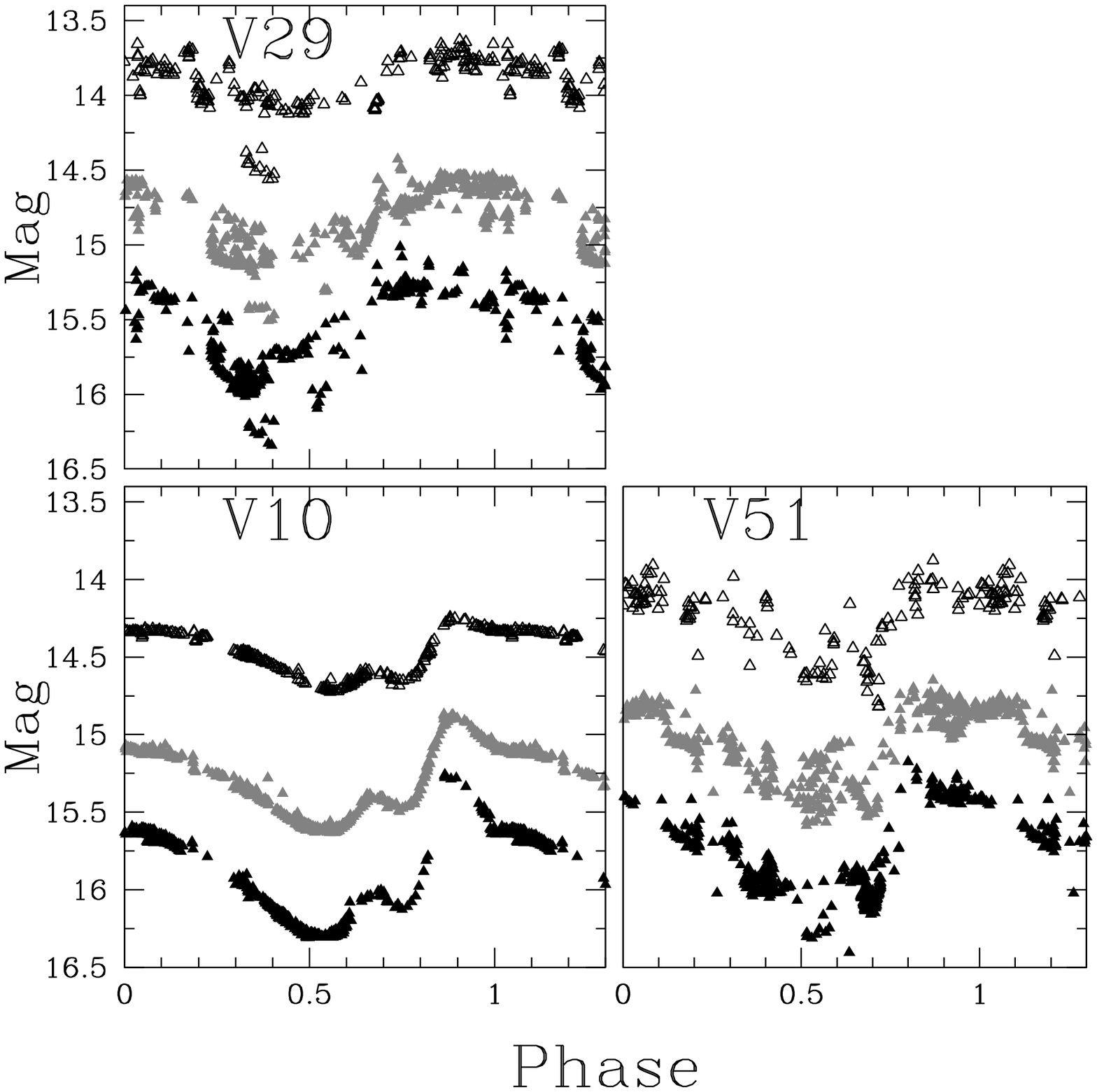}
\caption{Phased $BVI$ Light curves of the BL Her stars in NGC$\,$2808.
\label{lc_cw}}
\end{figure}

\begin{figure}[htb]  
\includegraphics[width=1\hsize]{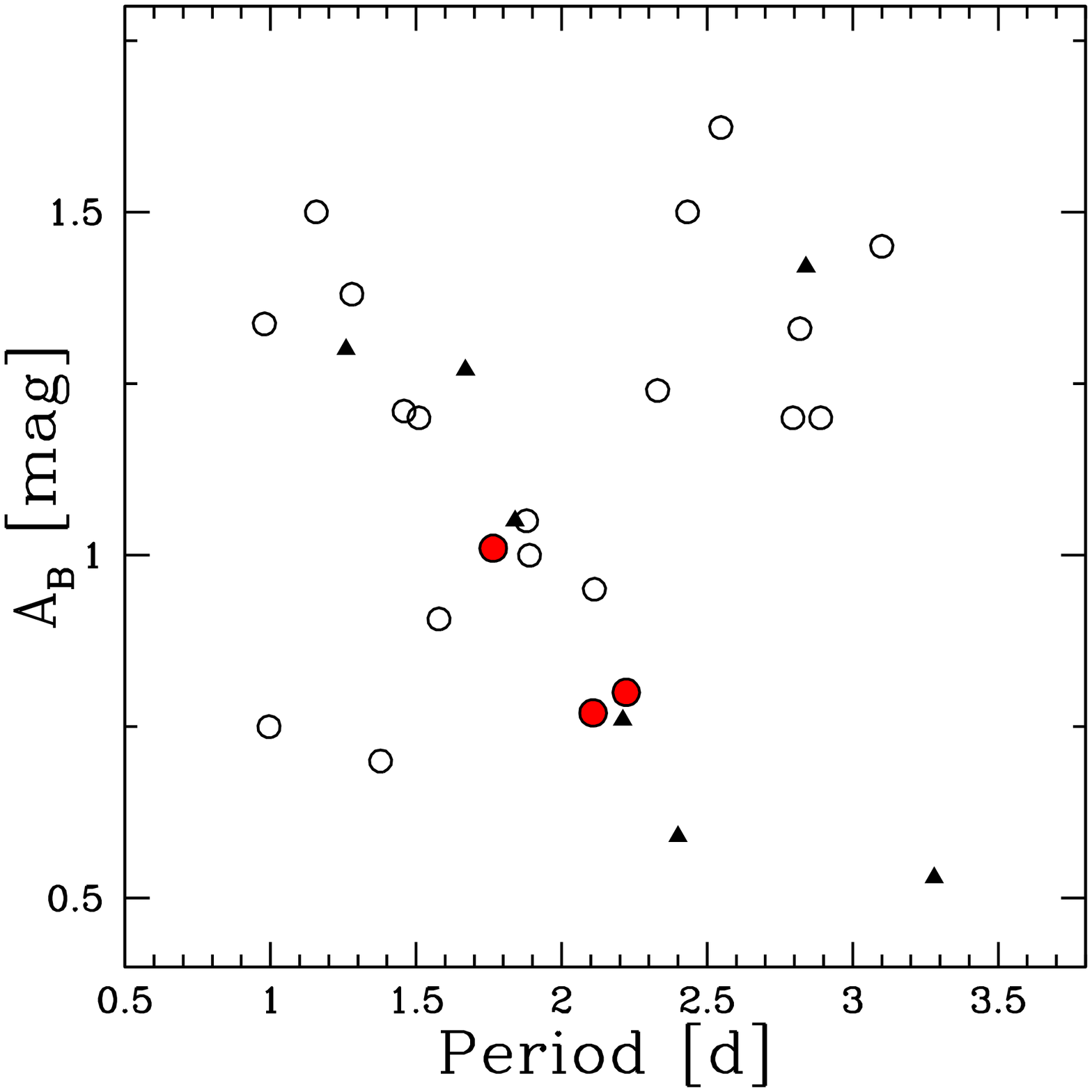}
\caption{Period-amplitude diagram of the BL Her stars in NGC$\,$2808.  The triangles
represent field BL Her stars from \citet{kwee68} whereas the open circles represent
GC BL Her stars from \citet{pritzl03}.
\label{blpa}}
\end{figure}

Population II Cepheids (P2Cs) with periods between about 1 and 8 days are usually 
designated BL Her variables.  These stars are more luminous than RRLs and 
in GCs are usually found in clusters with a blue HB morphology and few 
RRLs \citep[e.g.,][]{catelan09a}.  BL Her variables obey a well-defined period-luminosity 
relation and can be used as distance indicators \citep[e.g,][]{pritzl03, feast08, matsunaga09,
matsunaga11}.
We confirm V29 and V51 as BL Her variables and derive
updated periods and {\it BVI\/} magnitudes.  The light curves are shown
in Figure~\ref{lc_cw}.
Figure~\ref{blpa} shows the periods and $B$-amplitudes of the BL Her
stars in NGC$\,$2808 as compared to those in other GCs and the field.  
The data for the field BL Her stars comes from \citet{kwee68} and the data for the GCs
was assembled from \citet{pritzl03}.

Although V51 has properties indicative of a BL Her, there is more scatter in its phased light curve
than for the other two presented here.  The scatter may be caused by a 
secondary period, but BL Her stars are generally thought to be pulsating in one mode, presumably 
the fundamental.  As shown in \citet{schmidt05}, light curves of BL Her stars are 
not always very stable, and that may be the case here as well.  More observations would be 
especially helpful to investigate this star in more detail.

Using the absolute magnitude relations from \citet{pritzl03}, 

\begin{equation}
M_V = -1.64(\pm0.05) \log P + 0.05(\pm0.05),
\end{equation}

\begin{equation}
M_B = -1.23(\pm0.09) \log P + 0.31(\pm0.09),
\end{equation}

\begin{equation}
M_I = -2.03(\pm0.01) \log P - 0.36(\pm0.03),
\end{equation}

\noindent
the distance modulus to NGC$\,$2808 is found for all three passbands and listed in 
Table~\ref{bldist}.  The errors indicate the rms deviation of each relation.
We adopt $E(\hbox{\it B--V\/})$=0.17 mag (see \S4.2) 
and assume the canonical
Galactic coefficients of selective extinction, $R_V$=3.1, $R_B$=4.1 and $R_I$=1.8.
The mean distance modulus is $(m-M)_{0,BLHer}$= 14.99$\pm$0.06, where 0.06 is the error 
in the mean (and does not take into account the rms in the absolute magnitude relations).  
This is slightly lower than the \citet{harris96}
value of $(m-M)_{0}$= 15.06 (obtained using the same reddening).

\citet{pritzl03} calibrate these absolute magnitude relations using P2Cs from NGC$\,$6441
and NGC$\,$6388, clusters more metal-rich than NGC$\,$2808.  However, as shown by
\citet{mcnamara95} and \citet{pritzl03}, there is no convincing evidence that the
absolute magnitudes of these variables depend on $\rm [Fe/H]$.

\begin{table}
\begin{scriptsize}
\centering
\caption{Distance Determinations from BL Her variables in NGC$\,$2808}
\label{bldist}
\begin{tabular}{p{0.3in}p{0.6in}p{0.6in}p{0.6in}p{0.6in}} \\ 
\hline
Name & Period (d) & $\rm (m-M)_{0,B}$ & $\rm (m-M)_{0,V}$ & $\rm (m-M)_{0,I}$  \\ 
\hline
V10 & 1.76528 & 15.21$\pm$0.10 & 15.11$\pm$0.07  & 15.03$\pm$0.06 \\
V29 & 2.22211 & 14.94$\pm$0.10  & 14.83$\pm$0.07  & 14.68$\pm$0.06 \\
C51 & 2.10797 & 15.12$\pm$0.10  & 15.04$\pm$0.07  & 14.97$\pm$0.06 \\
\hline
\end{tabular}
\end{scriptsize}
\end{table}
\section{Other Variables}
V30 -- V30 was discovered by \citet{corwin04} and identified as an eclipsing binary star.  
Its light curve is shown in Figure~\ref{lc_v30}, and is a detached eclipsing binary.  
Although its $V$-light curve is relatively complete, more observations, especially in
the $B$ and $I$ passbands, would be helpful to determine the minima of the eclipse.  

C54 -- This is a new candidate variable star, an equal-light eclipsing binary 
located in the blue straggler region of the CMD (see Figure~\ref{rrcmd}).  
From a sample of 14 GCs and 20 blue straggler contact binaries, 
\citet{rucinski00} found that their frequency of occurrence is one such system 
(counted as one object) per 45 $\pm$ 10 blue stragglers.  As more of such
systems are being identified, the frequency of contact binaries among blue
stragglers can be used to constrain their currently unknown formation 
scenario \citep[e.g.,][]{mateo90, beccari08}.

\begin{figure}[htb]  
\includegraphics[width=1\hsize]{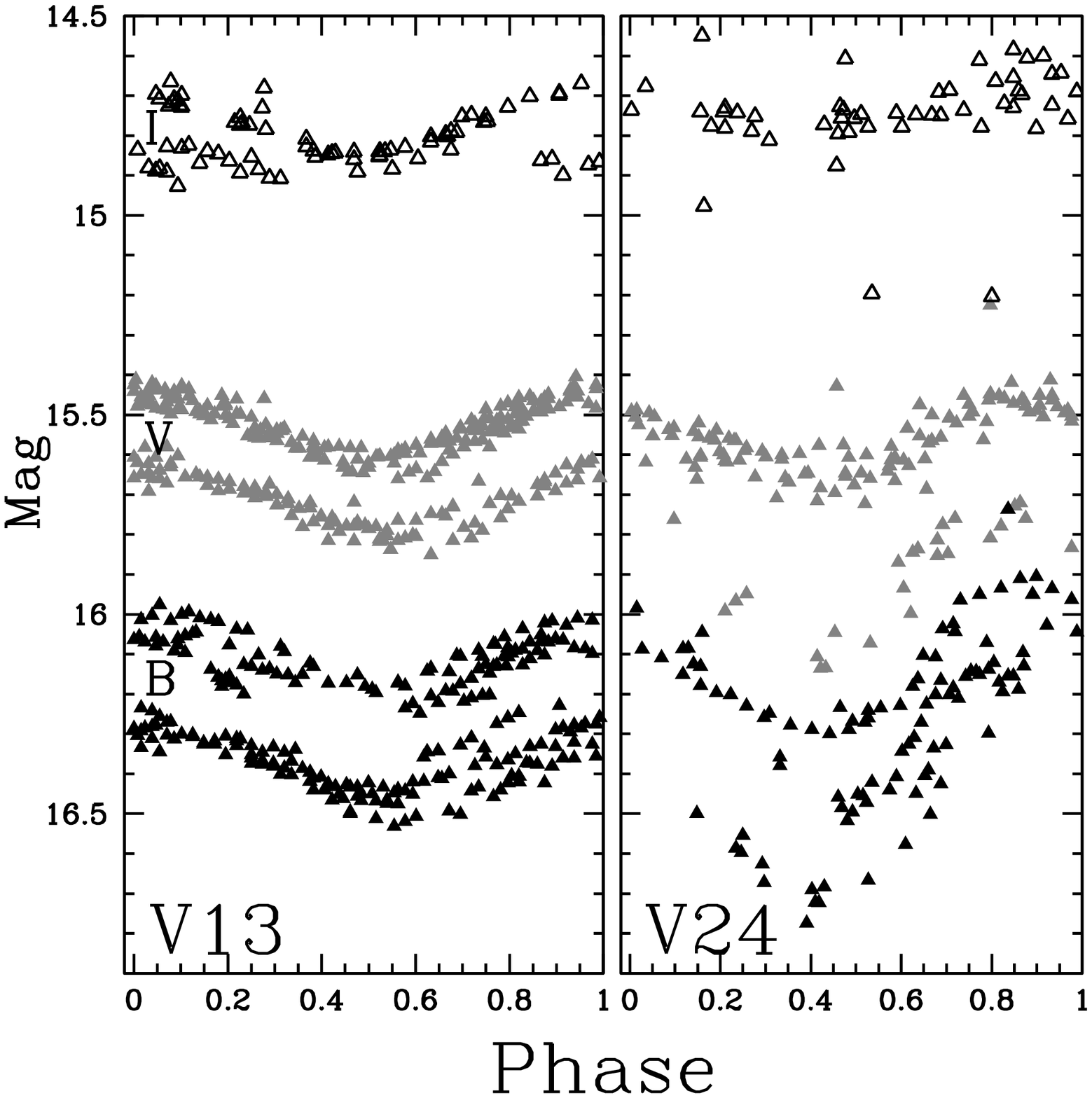}
\caption{Phased $BVI$ light curves of star V13 (right) and V24 (right) observed in NGC$\,$2808.  
\label{lc_v13v24}}
\end{figure}

V24 -- V24 was discovered by \citet{corwin04} and classified as an RR1.  Low-amplitude
variables are difficult to identify and classify \citep{kinman10}, and \citet{hoffman09}
point out that especially without color information and higher-precision photometry, the 
W Ursae Majoris type variables (W UMa) and RR1 variables cannot easily be differentiated.  
Our photometry of V24 suggests that this star is too bright to be an RR1 variable
belonging to the cluster, and therefore 
it may be a field W UMa or an ellipsoidal binary.  
Figure~\ref{lc_v13v24} shows that this star has a noisy, roughly sinusoidal light curve, characteristic of
many W UMa stars, which are known to have light curves with shapes and average brightness levels
that vary from one season to another \citep[e.g.,][]{vantveer91, rucinski02, borkovits05, hoffman06,
pribulla06}.  The orbital period found is in good agreement with that of W UMa systems 
\citep{stepien01, rucinski07}.  However, the amplitude is decreasing when going from $B$ to
$I$ (see Table~2), which is not a feature of W UMa stars.
This star could also be a foreground RR1 star.

\def\Min{${}^{\prime}$\llap{.}}
V24 is situated well within the 0\Min8 half-light radius of the cluster, so blending is always a possibility.
However, if blending has brightened the star by 0.75 mag, then the blending companion must be as bright 
as the variable.  From the HST thumbnail, such a large amount of blending is almost certainly not the case.
We therefore still believe this star is too bright to be an RRL.
On the other hand, the ratty light curve suggests that there could be some photometric problems, 
although we were unable to find a correlation between the magnitude offset and the seeing.  
The stars' chi and sharp indices are reasonable as compared to the other variable stars
presented here.


V13 --  V13 was discovered by \citet{corwin04} and tentatively classified as an 
RR1, although they noted that it might be RR2 (i.e., a second-overtone RRL) or 
a W UMa-type eclipsing binary.  As with V24, this star is too bright to be an RRL
belonging to the cluster, and
may be a field W UMa, a field ellipsoidal binary or a field RR1.  Its light curve is shown in Figure~\ref{lc_v13v24}.  
\def\min{${}^{\prime}$}
This star lies $\sim$7\min\ from the center,
and is one of the most distant variables from the cluster center. 
As there are no stars in its immediate vicinity, it is almost certainly not blended,
unless it is a real binary (separation $<<$0.5 arcsec, 
in a region where the mean separation between stars is tens of arcseconds). 

\subsection{Are V13 and V24 cluster members?}
As V13 and V24 lie well within a 8\min\ radius from the cluster center and
have colors and magnitudes within the range of 15.4 $> V >$ 16.5 
and 0.7$>$$\hbox{\it (B--I)\/}$$>$1.5, this distance, magnitude and color range
is used to obtain a rough estimate on the likelihood of finding such field variable stars in
the NGC$\,$2808 vicinity. 
Our observations include $\sim$140 stars with the above specified parameters,
two-thirds of those stars lying closer than
2\min\ from the cluster center.  This is similar to the number of disk stars predicted
from simulated star counts from the Besan\c{c}on model \citep{robin03}.
W UMa binaries occur in our solar neighborhood with an apparent frequency of 
one binary among about 500 single stars in the Galactic disk \citep{rucinski06}.  
Therefore, although unlikely, it is tenable to find a field W UMa with a magnitude
and color similar to V13 and V24.  However, it is suspicious for {\it two\/}
W UMa stars with such similar colors and magnitudes to be 
found in the field of NGC$\,$2808.  
V13 and V24 have $V$ magnitudes that are $\sim$0.5-0.8~mag brighter than
the RR Lyrae stars and colors that place them within the red and blue edges 
of the RR Lyrae instability strip (see Figure~\ref{rrcmd}).  These features
are reminiscent of Anomalous Cepheids (ACs).  In a system such as a dwarf 
spheroidal galaxy, ACs are $\sim$0.5 mag brighter than 
RR Lyrae stars at periods of $\sim$0.3 d, and the more luminous ACs with 
periods $\sim$2.0 days are $\sim$2.0 mag brighter than RRLs \citep{cox86, bono97b, marconi04}.  

\begin{figure}[htb]  
\includegraphics[width=1\hsize]{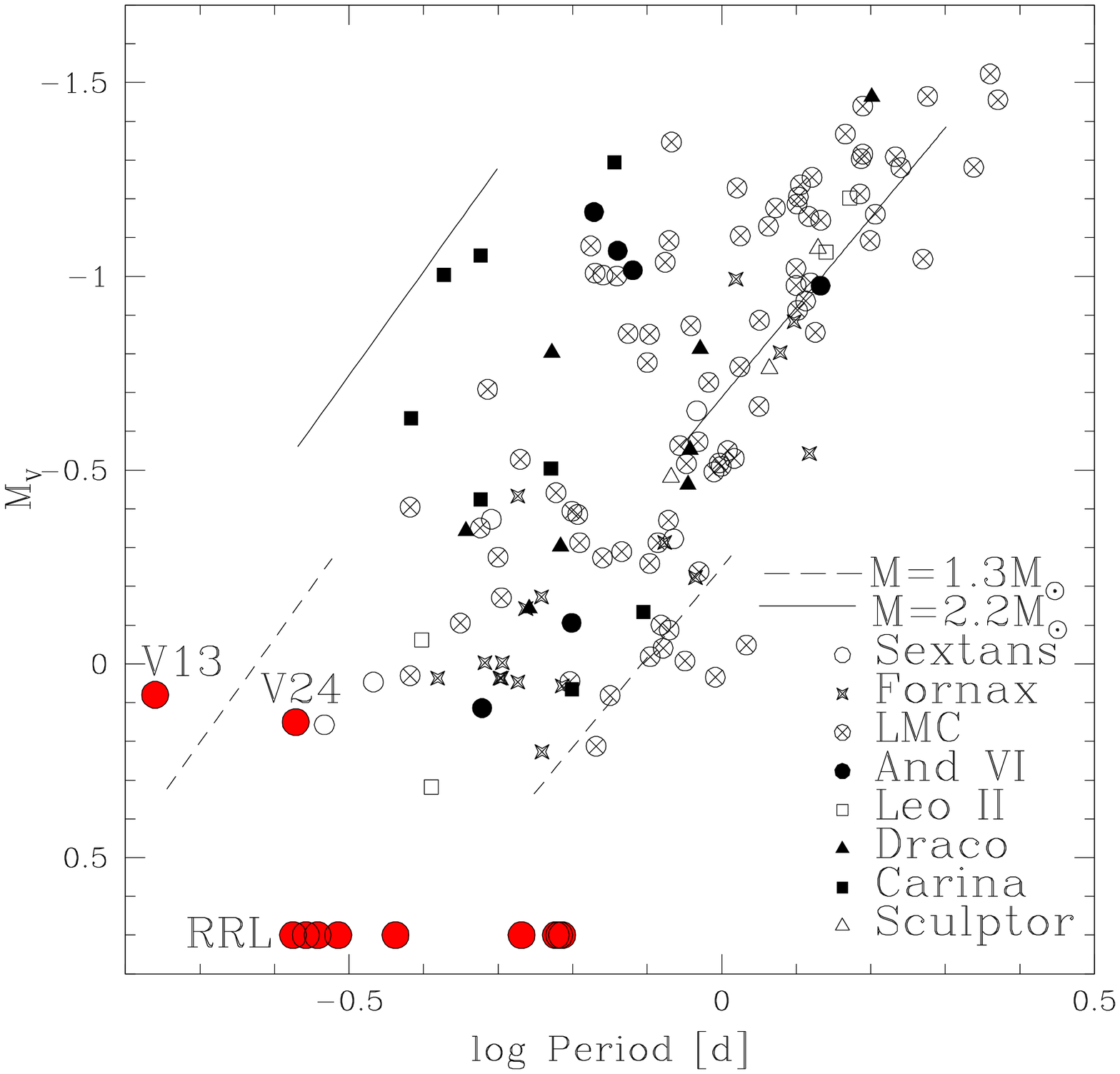}
\caption{The observed Anomalous Cepheids in dwarf spheroidal galaxies as compared with the
predicted edges of the instability strip in the $M_V$-log$P$ plane, from equations~5 and 6
of \citet{marconi04}.  The dashed line indicates the predicted IS at 1.3$M_\odot$ and the solid
line at 2.2$M_\odot$. To account for the occurrence 
of first-overtone pulsators, the blue limit is shifted by $\delta$log $P$ = $-$0.13 d.
The NGC$\,$2808 variables V13 and V24 are also shown, assuming they are located
at the distance of the cluster.
\label{ac_is}}
\end{figure}

\begin{figure}[htb]  
\includegraphics[width=1\hsize]{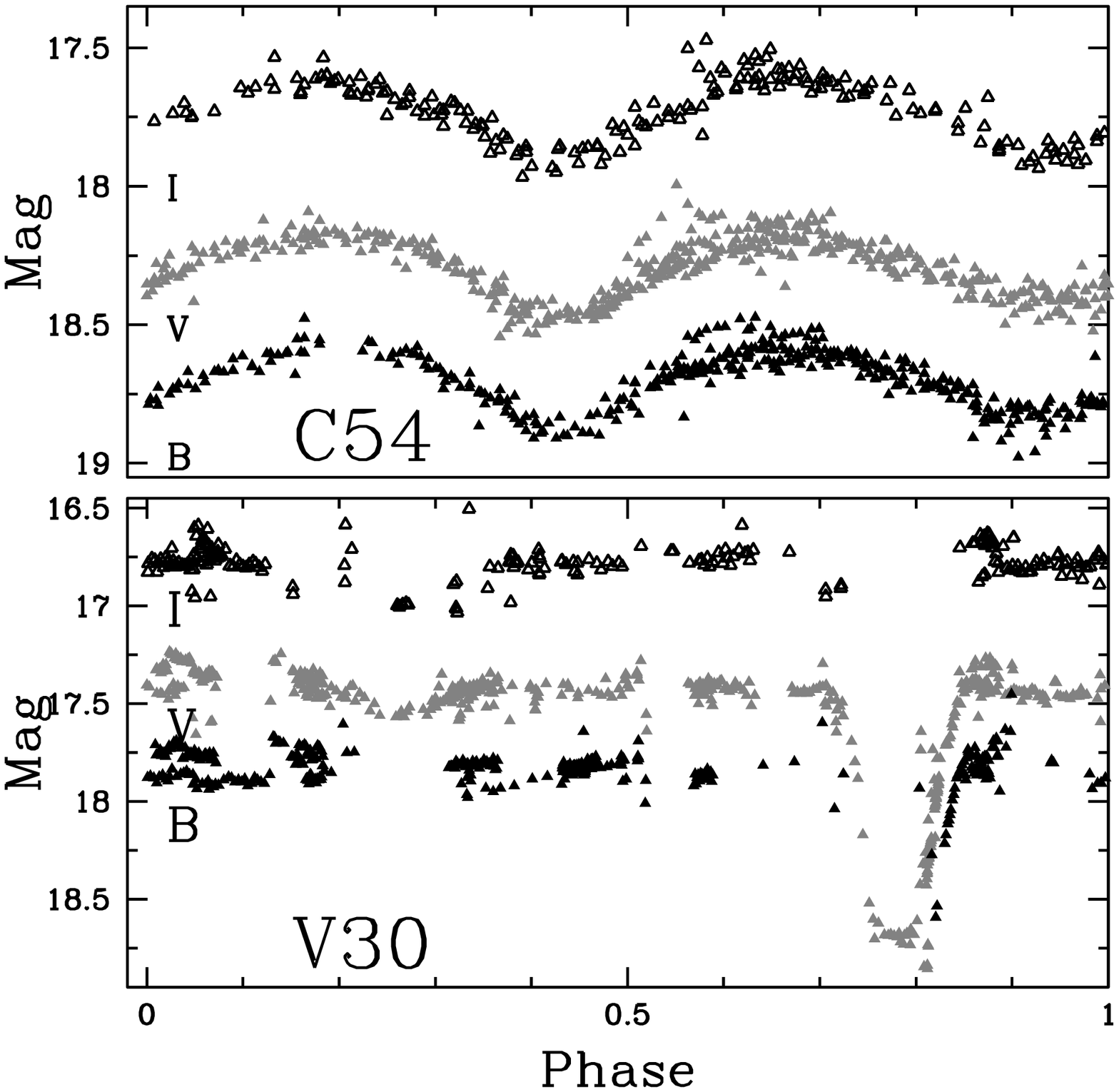}
\caption{Phased $BVI$ light curves of the eclipsing binary V30 observed in NGC$\,$2808.
\label{lc_v30}}
\end{figure}

If V13 and V24 are cluster members and assuming that the RRLs
have absolute magnitudes of $M_{V,RRL}$=0.70 (see \S4.1), the absolute magnitudes of
these stars are $M_{V}$$\sim$0.2 mag.  Figure~\ref{ac_is}, 
taken from \citet{marconi04}, shows the data for well-recognized ACs in dwarf 
spheroidal galaxies using the distance moduli provided by the various authors 
(see Marconi et~al. 2004, for references). V13 and V24 have shorter periods and smaller amplitudes
than the majority of the ACs in these galaxies, although such ACs may also
be the most likely missed when looking at the comparatively great distances to the dwarf spheroidal galaxies.
%
\citet{chiosi93} derive mass-period-luminosity-color relations in the $BV$ and $VI$ 
passbands for Cepheids, resulting from pulsation theory alone.  From their theoretical 
calculations, we derive log $\rm T_{eff}$ of $\sim$3.73 and $M$ of $\sim$1.5$M_{\sun}$ for V13 
and log $\rm T_{eff}$ of $\sim$3.73 and $M$ of $\sim$1.0$M_{\sun}$ for V24, in good agreement 
with those found for ACs \citep[e.g.,][]{bono97b, marconi04}.  However, as clearly seen, 
V13 has a period that is quite different from an AC, placing this star outside the limits of the
instability strip for its luminosity.  
Although ACs have been observed in many nearby Local Group dwarf galaxies, 
independent of their morphological type, only one or two ACs are known in Galactic
GCs \citep[V19 in NGC$\,$5466, and possibly V7 in NGC$\,$6341,][]{zinn76,matsunaga06}.  
We note that the light curve of V13 is similar to the AC in NGC$\,$5466, with 
a brighter and a fainter sinusoidal curve varying symmetrically with each other.

V13 and V24 may also be BL Her variables, especially since BL Her stars 
tend to reside in GCs with extended blue tails on the HB, like 
NGC$\,$2808 \citep[e.g.,][]{harris85,sweigart76,bono97c,wallerstein02, catelan09}.
However, again the periods of V13 and V24 are inconsistent with BL Her variability 
($\sim$0.2 days versus the typical BL Her period range of 1 - 3 days).  First-overtone 
BL Her stars are expected to have shorter periods and smaller amplitudes, but 
FO models are seen to ``snuggle right up" to the fundamental 
pulsators \citep{buchler92, moskalik93, buchler94}, so periods 
in the range of $\sim$0.2 d may still be too short for FO BL Her 
variables.  Quantitatively it is found that the 
blue edge for BL Her first-overtone pulsation is only $\sim$100 K from the fundamental 
one, producing a very narrow region of FO-only pulsation \citep{buchler92, buchler94}.  
This result is also supported by updated BL Her models from \citet{dicriscienzo08}.  
Lastly, first-overtone BL Her variables are thought to be bluer than their fundamental-mode 
counterparts \citep{nemec94} and V13 and V24 have colors similar to their 
fundamental-mode BL Her counterparts.  

We are unable to find a secondary period for V13 and V24 from our present data, 
although we cannot rule out the possibility.  \citet{soszynski08} detected the largest 
sample of P2Cs and ACs outside the Galaxy, finding a surprisingly large group of 
P2Cs with low-amplitude secondary periodicities.  More observations would be 
desirable for a more complete frequency analysis for these stars.
Spectroscopy would be useful to determine their radial velocities and 
therefore the probability of cluster membership.  Spectra may also be able to disentangle 
whether these stars are varying because of an orbiting companion or because
of intrinsic effects such as pulsation.

\section{Discussion and Conclusions}
We have presented the first calibrated {\it BVI\/} photometry for 11 of the 18 previously 
classified RR Lyrae variables in the multimodal-HB, split-MS globular cluster NGC$\,$2808.  
Photometry of the other seven 
was hampered by the extreme crowding 
present within the cluster, and space-based observations are likely needed for the determination
of calibrated light curves for these stars.  At least two of these stars are too bright to be RR Lyrae 
variables, leading to a revision of the probable number of RR Lyrae stars to 16 and a specific RR Lyrae 
fraction of $S_{RR}$=2.8.  Here, $S_{RR} = N_{RR} \times 10^{0.4(7.5+M_V)}$, 
where $M_V$=$-$9.39 \citep[][2010 update]{harris96} is the cluster's integrated 
absolute magnitude in $V$.  

We have derived pulsational parameters for all the RR Lyrae variables.
Five of the 16 have periods consistent with first-overtone pulsators,
and eleven have periods consistent with fundamental-mode pulsators, although
the light-curve fit to our data for some of the RR0 stars is noisy (i.e., V17, V19, V21, V25).  Our
adopted periods for these stars seem to be a reasonable compromise.
The mean periods of the RR0 and RR1 variables are $\rm <P_0>$ = 0.56$\pm$0.01 days 
and $\rm <P_1>$= 0.30$\pm$0.02 days, respectively, and the number ratio of the RR1-type 
variables to the total number of the RRL-type variables is $N_1/N_{RR}$ = 0.3. 
Using the period-amplitude diagram, the Oosterhoff type of NGC$\,$2808 is that of an OoI cluster,
which is expected considering the RR Lyrae mean periods as well as the fairly high metallicity of NGC$\,$2808 
($\rm [Fe/H]$ $\sim$ $-$1.14 dex; Harris 1996, 2010 edition). 
On the other hand, the number ratio of RR1 to total RR Lyrae stars is on the high end of typical OoI-type clusters.

The mean magnitude for the RRLs is $\rm m_{V,RR}$ = 16.21 $\pm$ 0.04 mag,
and the reddening from their minimum light colors was determined to 
be  E(\hbox{\it B--V\/}) = 0.17$\pm$0.02 mag, in 
agreement with previous studies \citep[e.g.,][]{walker99, bedin00}.  
Using the recent recalibration of the RR Lyrae luminosity scale by \citet{catelancortes08}, 
the RR Lyrae variables have absolute magnitudes of
$M_V$=0.70$\pm$0.13 mag.  This leads to an RR Lyrae distance of 
$(m-M)_{V,RRL}$=15.57$\pm$0.13 and adopting E(\hbox{\it B--V\/})=0.17 mag, 
$(m-M)_{0,RRL}$=15.04$\pm$0.13 mag. 

From the comparison of theoretical ZAHB sequences with observations, we find
that most RRLs have a mass of M=0.57-0.60$M_\sun$ and that the
the RRL IS is well matched by models with standard He values ($Y$=0.248 in this case).
That the NGC$\,$2808 RRL are not He enhanced is also supported by their derived 
$V$-amplitudes and periods.  Their pulsational properties are similar to the RRL 
in M3, a cluster in which any helium enhancement
is very likely less than 0.01 along the HB \citep{catelan09}. 
In fact, all of the RRL have OoI-type characteristics and do 
not show large period shifts (see Figure~\ref{PA}), such as are expected 
for RRL with high luminosities indicative of 
He-enrichment \citep{sweigart98, pritzl02, sollima06, marconi11, kunder12}.

Still we are hesitant to associate the RR Lyrae stars with the red HB.  
This is largely because it is unclear what enhancement in the He abundance
to expect for a given level of $\rm [O/Fe]$ depletion and $\rm [Na/Fe]$, $\rm [Al/Fe]$
enhancement. (see Catelan 2012, for a recent discussion).
For example, \citet{marcolini09} show that there is a wide range of possible He values for the same 
predicted $\rm [O/Fe]$ depletion level -- including some very small He enhancement, 
of order $\sim$0.01, for $\rm [O/Fe]$ $<$ $-$0.5 (see their Fig.~16).  It is therefore not 
impossible that the RR Lyrae stars are indeed associated with the hotter HB component 
of the cluster, but with a level of He enhancement that is just too small to manifest itself 
clearly in our data.

Except for one, all of the RR Lyrae variables studied, both the fundamental and first-overtone, show 
evidence of light curve modulation.  This is abnormal and may be partially due to inaccuracies
in our ground based photometry and/or the Blazhko effect.  The precision of our photometry
is sometimes 0.05 mag, for the most crowded RR Lyrae stars, making it difficult to confirm or
refute Blazhkocity in the variables.  Space based observations,
image subtraction software and more photometry would help shed light on which stars are Blazhko 
stars and which are affected by crowding/blending issues, large period
change rates, double-mode pulsation or the period doubling phenomenon.
If the variables in NGC$\,$2808 are confirmed as Blazhko stars 
the unusually high Blazhko ratio would be similar to that of NGC$\,$5024 (M53), a GC hosting
one of the largest percentages of Blazhko variables, 36\% and 66\% of the
total population of RR0 and RR1 stars in the cluster, respectively \citep{arellano12}.  
However, unlike M53, NGC 2808 is relatively metal-rich.  A high percentage
of Blazhko stars in a metal-rich system would be in contrast to the suggestion of a 
correlation between the occurrence of the Blazhko effect and metallicity.

Our observations were insufficient to discern the nature of two variable stars, V13 and V24. 
These two stars are $\sim$0.8 mag brighter than the RR Lyrae variables, have colors
that place them near the red edge of the RRL IS, and have noisy light curves.
Their periods of $\sim$0.2 days make it difficult to classify these stars as either ACs or BL Her stars.
However, statistically it is curious to find two field W UMa variables or ellipsoidal binaries with
similar colors and magnitudes to not only each other, but also to the NGC$\,$2808 RRL and BL Her stars.  

NGC$\,$2808 has been tentatively associated with the Canis Major dwarf spheroidal galaxy.
The RRL pulsational properties are somewhat unusual compared 
with bona-fide Galactic GCs, but unlike what is typically seen in the Galaxy's dwarf satellites
(e.g., Catelan 2009a, and references therein), there is no evidence that NGC$\,$2808 is an
Oosterhoff-intermediate GC.  
The reddening and distance modulus derived in this work are similar to those 
previously accepted for the cluster \citep{harris96}, and so the conclusions reached by 
\citet{crane03} and \citet{forbes04} regarding its possible association with this dwarf galaxy are not 
significantly affected by these new estimates.  Although NGC$\,$2808 is not particularly 
RR Lyrae-rich, it does harbor RR Lyrae variables, unlike the field stars located in Canis 
Major \citep{kinman04, mateu09}.  If V13 and V24 are shown to be ACs, this would
suggest an extragalactic origin for NGC$\,$2808, as ACs are found mainly in
dwarf spheroidals and are very rare, though not unprecedented, in Galactic GCs.

Two new candidate SX Phe stars and one new candidate eclipsing binary in the blue
straggler region of the CMD are found in the field of NGC$\,$2808.  The SX Phe and BL Her stars have 
typical pulsational properties compared to other Milky Way GCs.  From the three BL Her stars 
in NGC$\,$2808, the distance to the cluster is  $(m-M)_{V,BLHer}$= 15.50$\pm$0.12 and from the two 
SX Phe stars, $(m-M)_{V,SXPhe}$= 15.58$\pm$0.12 is found,
in good agreement with the distance determined by the RRLs.

\acknowledgments
We are grateful to Giuseppe Bono for helpful discussions regarding V13 and V24.
M.C. and P.A. are supported by the Chilean Ministry for the 
Economy, Development, and Tourism's Programa Iniciativa Cient\'{i}fica  
Milenio through grant P07-021-F, awarded to The Milky Way Millennium  
Nucleus; by the BASAL Center for Astrophysics and Associated Technologies  
(PFB-06); by Proyecto Fondecyt Regular \#1110326; and by Proyecto Anillo  
ACT-86. 

This work has made use of BaSTI web tools.

\clearpage

\end{document}